\documentclass[sigconf]{acmart}
\usepackage[english]{babel}
\usepackage{blindtext}

\usepackage{booktabs} 
\usepackage{array}   
\usepackage{tabularx}
\usepackage{hyperref}
\usepackage{url}
\usepackage{xurl}
\usepackage{listings}

\usepackage{listings}
\usepackage{xcolor}
\usepackage{tcolorbox}
\tcbuselibrary{breakable}
\usepackage{enumitem}
\lstdefinestyle{jsonstyle}{
    backgroundcolor=\color{gray!10},
    basicstyle=\ttfamily\small,
    breaklines=true,
    frame=single,
    tabsize=2,
    showstringspaces=false,
    escapeinside={(*@}{@*)}
}
\usepackage{minted}
\renewcommand\footnotetextcopyrightpermission[1]{} 
\setcopyright{none}

\newcommand{\parab}[1]{\vspace{0.03in}\noindent\textbf{#1}}

\acmConference[]{}{}{}
\setcopyright{none}
\renewcommand\footnotetextcopyrightpermission[1]{}

\begin{document}
\title[Airavat]{Airavat: An Agentic Framework for \\ Internet Measurement}

\author{Alagappan Ramanathan}
\orcid{0009-0003-3293-1790}
\affiliation{%
   \institution{University of California, Irvine}
}
\author{Eunju Kang}
\orcid{0009-0005-3591-5758}
\affiliation{%
    \institution{University of California, Irvine}
}
\author{Dongsu Han}
\orcid{0000-0001-6922-7244}
\affiliation{%
  \institution{KAIST}
}
\author{Sangeetha Abdu Jyothi}
\orcid{0009-0000-0503-4478}
\affiliation{%
   \institution{University of California, Irvine}
}
\thanks{Sangeetha Abdu Jyothi holds concurrent appointments at UC Irvine and Amazon. This publication describes work performed at UC Irvine and is not associated with Amazon}

\renewcommand{\shortauthors}{Alagappan Ramanathan, Eunju Kang, Dongsu Han, Sangeetha Abdu Jyothi}

\begin{abstract}
    Internet measurement faces twin challenges: complex analyses require expert-level orchestration of tools, yet even syntactically correct implementations can have methodological flaws and can be difficult to verify. 
Democratizing measurement capabilities thus demands automating both workflow generation \textit{and} verification against methodological standards established through decades of research.

We present Airavat, the first agentic framework for Internet measurement workflow generation with systematic verification and validation. Airavat coordinates a set of agents mirroring expert reasoning: three agents handle problem decomposition, solution design, and code implementation, with assistance from a registry of existing tools. Two specialized engines ensure methodological correctness: a Verification Engine evaluates workflows against a knowledge graph encoding five decades of measurement research, while a Validation Engine identifies appropriate validation techniques grounded in established methodologies. Through four Internet measurement case studies, we demonstrate that Airavat (i) generates workflows matching expert-level solutions, (ii) makes sound architectural decisions, (iii) addresses novel problems without ground truth, and (iv) identifies methodological flaws missed by standard execution-based testing.

\end{abstract}

\maketitle

\section{Introduction}

Agentic AI systems leveraging Large Language Models (LLMs) have demonstrated remarkable capabilities across diverse domains, from code generation to scientific reasoning~\cite{2024codellamaopenfoundation, wei2023chainofthoughtpromptingelicitsreasoning, Wang_2024}. Their ability to decompose complex problems, explore solution spaces, and synthesize executable implementations positions them as powerful tools for automating sophisticated analytical workflows. However, applying agentic systems to Internet measurement research faces two fundamental challenges that limit their practical deployment.

\textit{First}, Internet measurement analyses require expert-level orchestration of multiple specialized tools---BGP analyzers~\cite{routeviews, riperis, bgptools, bgpstream, bgpdetecting, bgpstate, bgpnext, albgp}, traceroute processors~\cite{tracerouteinternet, traceroutemao, paristraceroute, traceroutemultilevel}, topology mappers~\cite{topologyzoo, nautilus, topologycollecting}, and performance monitors ~\cite{perfsurvey, gatechIODA, netblocks, xaminer, cloudflare}---each with unique interfaces, data formats, and domain knowledge requirements. When researchers need to understand routing behavior, infrastructure dependencies, or performance anomalies, they must manually integrate different measurement systems through custom solutions.

Recent events highlight this challenge's practical impact. The AAE-1 cable cuts~\cite{multiple_providers_aae1_outage} and FALCON cable failure~\cite{falcon_outage} caused widespread outages, requiring rapid development of workflows integrating cable mapping, BGP analysis, and traffic flow assessment. Similar challenges arise regularly: understanding CDN performance degradation requires correlating traceroute data with BGP changes~\cite{latlong, cdnmoving, cdnserver}; investigating security incidents demands integrating multiple measurement perspectives. While recent measurement frameworks~\cite{nautilus,xaminer} offer powerful capabilities, these tools operate in isolation and require specialized knowledge. Experts must spend days developing measurement workflows before analysis can begin, creating a substantial barrier: the ability to compose advanced measurement workflows requires specialized domain experience, limiting such capabilities to a small community of experts.

\textit{Second}, unlike agentic systems for system optimization whose evolution is guided by well-defined performance goals, verifying and validating measurement workflows is inherently difficult. Agentic systems may generate executable code for Internet measurement with methodological flaws that corrupt analytical results while appearing to function properly.  Traditional software testing approaches---checking code syntax, validating output formats, and ensuring execution completes---prove insufficient for measurement workflows, where correctness depends on tacit domain knowledge of data artifacts, appropriate preprocessing steps, and methodological precedents established over decades of research. For instance, an agent might correctly implement longest-prefix-match operations yet fail to filter routing table artifacts, producing meaningless results despite successful execution. 

We take an alternative view. What if network operators could ask high-level questions in natural language and receive executable measurement workflows in minutes? What if researchers could compose Internet measurement tools without specialized training in each framework while ensuring generated solutions meet established quality standards? 

We present \textit{Airavat}, the first domain-specific agentic framework for Internet 
measurement workflow generation with systematic verification and validation capabilities. 
Airavat addresses the dual challenges by automating workflow generation and verification, while also supporting human-in-the-loop oversight to ensure methodological rigor. 
Our key insight is that measurement workflow development follows predictable patterns 
decomposed into distinct phases: problem analysis, solution design, and implementation. Airavat executes these phases through three specialized agents---QueryMind, 
WorkflowScout, and SolutionWeaver---operating on a curated Registry of 
measurement tools. To improve trust, Airavat adds two specialized engines that systematically 
verify and validate generated workflows against established measurement methodologies, 
ensuring methodological correctness before deployment. Users express 
measurement goals in natural language, and the system generates executable measurement 
solutions that provide complete workflows or serve as foundations for expert refinement. 

To demonstrate Airavat's capabilities, we present four distinct Internet measurement case studies. Our evaluation demonstrates Airavat's ability to (i) independently generate workflows that produce analytical outputs similar to expert-designed solutions with the relevant solutions removed from the Registry and the Knowledge Graph (\S~\ref{sec:esr},~\ref{sec:cs2}), (ii) orchestrate complex analysis across multiple measurement frameworks with significant integration complexity (\S~\ref{subsec:cascading}), and (iii) identify critical methodological flaws missed by standard execution-based testing through systematic verification (\S~\ref{sec:verify-results}), and (iv) synthesize validation methods for generated workflows guided by prior literature (\S~\ref{sec:validate-results}).

In summary, we make the following contributions.

\begin{itemize}[leftmargin=*]

\item We present Airavat, the first domain-specific agentic framework that translates high-level Internet measurement queries into executable workflows.
\item We introduce a \textit{verification engine} in Airavat that evaluates generated workflows against five decades of Internet measurement research encoded in a structured knowledge graph. Unlike execution-based testing, our approach detects methodological flaws that silently corrupt results.
\item  We design a \textit{validation engine} that discovers, adapts, and implements appropriate validation techniques from prior literature, enabling measurement workflows to be checked using alternative methods.
\item Through extensive Internet measurement case studies, including cross-layer infrastructure resilience and IP allocation analysis, we demonstrate the capabilities of Airavat.
\end{itemize}

\begin{figure*}
    \centering
    \includegraphics[width=0.8\linewidth]{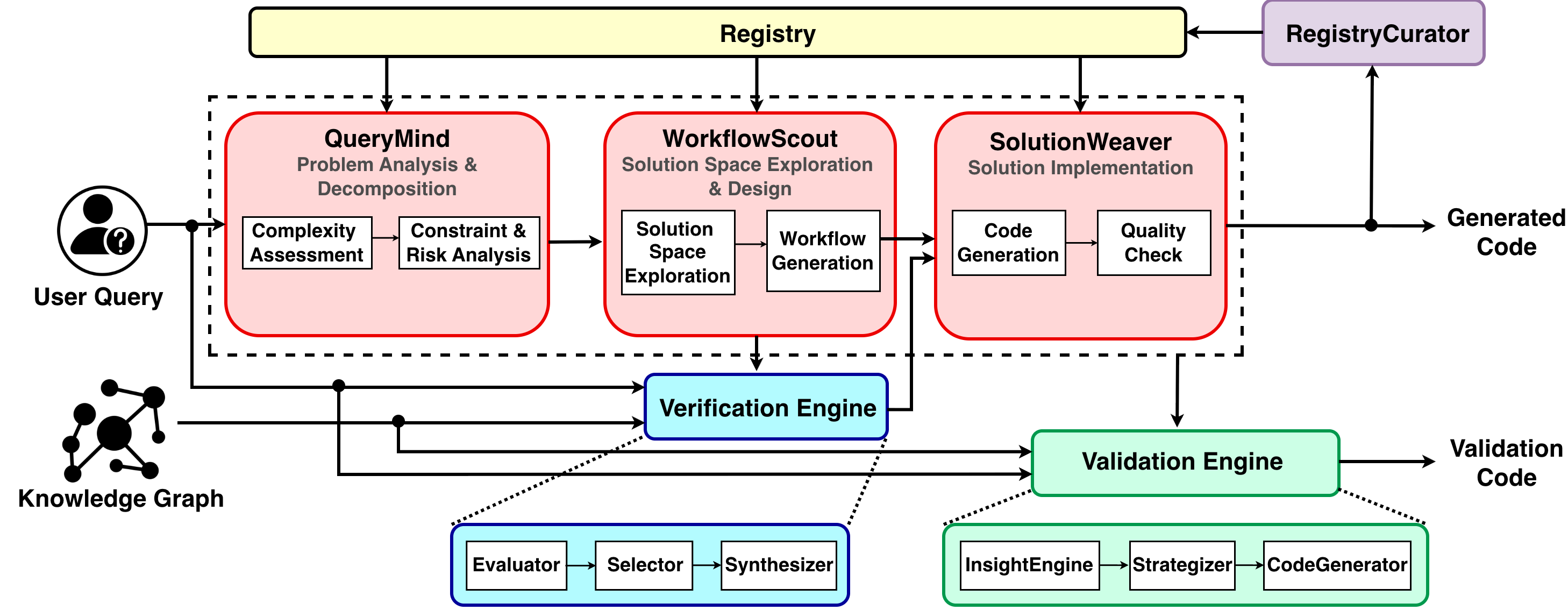}
    \caption{\small Airavat’s architecture comprising a multi-agent workflow generation pipeline, verification engine, validation engine, knowledge graph, and registry.}
    \label{fig:design_diagram}
\end{figure*}

\section{Airavat Design Overview}

Building on our vision for an agentic framework for Internet measurement~\cite{arachnet}, we present Airavat, which tackles the fundamental challenge of generating trustworthy measurement workflows from natural language queries. While our prior work demonstrated workflow generation capabilities, Airavat addresses additional core problems essential for trustworthiness. First, workflow design requires systematically decomposing complex measurement tasks into executable steps. Second, workflow verification and validation must ensure that generated workflows are scientifically rigorous and methodologically sound.

To address these challenges, Airavat integrates specialized subsystems shown in Figure~\ref{fig:design_diagram}. The \textit{Multi-Agent Workflow Generation} pipeline emulates expert reasoning, with specialized agents handling problem decomposition, solution design, and implementation. The \textit{Registry}, a catalog of measurement tools, assists this process. The \textit{RegistryCurator} enables evolution of the registry by extracting proven patterns from generated workflows and codifying them as reusable capabilities. To enable workflow verification, we construct the \textit{Knowledge Graph}, which encodes measurement literature as semantic relationships and validation patterns that enable literature-grounded quality assessment.  The \textit{Verification Engine} evaluates workflow quality through a literature-grounded assessment and addresses identified deficiencies. The \textit{Validation Engine} generates executable validation code by identifying applicable approaches from the literature and adapting them to specific problems. 

\noindent \textbf{Scope.} Airavat focuses on workflow composition and code generation for measurement analysis—weaving together existing measurement tools and data sources to generate executable measurement solutions that provide complete workflows or serve as foundations for expert refinement. Out of scope are distributed measurement collection itself, tool implementation improvements, novel data acquisition, and improving individual measurement tools.

\section{Workflow Generation}

In this section, we detail of the key components enabling workflow generation.

\subsection{Registry: Tool Base}

To reason about workflow composition, agents need structured knowledge about available measurement capabilities. The Registry is a manually curated catalog describing what measurement tools can do, not how they do it. This abstraction emerged from early experiments where exposing entire codebases overwhelmed agents, causing them to miss key capabilities. Each registry entry specifies an open-source tool's capabilities, its required inputs, expected outputs, and operational constraints in a standardized format. This abstraction scales linearly with available tools rather than with codebase complexity—a framework with 10,000 lines of code contributes a single registry entry. The standardized specification enables agents to evaluate tool applicability, understand integration requirements, and compose multi-tool workflows without framework-specific knowledge. We bootstrap the Registry with a set of manually curated tools. The Registry evolves organically as the RegistryCurator (\S~\ref{subsec:curator}) identifies reusable patterns from successful workflows and proposes new entries, though all additions undergo manual validation to maintain quality standards.

\subsection{QueryMind: Problem Decomposition}
The QueryMind Agent transforms user queries into structured problem representations by decomposing them into sub-problems, dependencies, and constraints. This agent solves a specific problem: natural language queries contain hidden complexity and implicit assumptions that must be made explicit before solution design can proceed. Decomposing "measure CDN performance" into latency analysis across regions, cache behavior evaluation, and temporal consistency checking enables subsequent agents to design targeted solutions rather than over-generalized workflows.

We use a task-agnostic prompt in QueryMind to systematically decompose measurement queries into manageable subproblems by examining five dimensions of complexity: temporal (evolution over time), spatial (geographic/network regions), causal (primary/secondary/tertiary effects), stakeholders (multiple perspectives), and data (complementary sources). It assigns a complexity score (0-5) that guides WorkflowScout's exploration strategy. The agent prioritizes early constraint evaluation: even a theoretically excellent workflow is infeasible if the required data or technical infrastructure is unavailable. It then defines success criteria to prevent under-analysis and over-engineering, and maps sub-problems to relevant Registry functions, providing the WorkflowScout Agent with focused guidance (detailed output schema in Appendix Figure~\ref{fig:querymind_schema}).

\subsection{WorkflowScout: Workflow Design}

The WorkflowScout Agent converts structured sub-problems into concrete solution architectures. This separation from implementation is essential: solution design requires exploring multiple competing approaches and evaluating trade-offs, a fundamentally different task from writing executable code. A monolithic approach either skips exploration to focus on implementation, or produces verbose code exploration that obscures the actual solution architecture.

WorkflowScout employs an adaptive exploration strategy that scales to problem complexity. Simple queries (complexity score 0-1 generated by QueryMind) receive direct solutions since alternatives provide minimal benefit. Moderate complexity (2-3) triggers a primary approach plus 1-2 complementary alternatives targeting missed insights. Complex problems (4-5) require three or more approaches from different analytical perspectives. 

This strategy reflects measurement research practice: stronger conclusions come from multiple independent methods rather than optimizing a single approach. For high-risk core components (steps that address primary success criteria or critical data transformations), the agent designs alternative approaches under different assumptions and failure modes, ensuring backup options if any approach fails. For method selection, the agent employs ensemble approaches over parameter tuning, combining multiple complementary methods using four composition patterns (detailed in Appendix~\ref{sec:patterns}). WorkflowScout produces a comprehensive design specification for the SolutionWeaver Agent.

\subsection{SolutionWeaver: Code Implementation}

The SolutionWeaver converts workflow designs into executable code that integrates heterogeneous measurement tools. Since independently designed frameworks may use diverse data representations, the agent implements format translation using registry specifications to ensure seamless data flow. A critical challenge is that LLMs produce plausible but non-executable code. Airavat addresses this through strict execution requirements, ensuring generated code processes real measurement data with complete implementations. Early experiments revealed systematic failure modes, such as synthetic data generation, placeholder functions, and oversimplified operations, prompting quality assurance through validation checkpoints embedded throughout generation. This produces code requiring minimal manual correction. The agent also documents functions with reusability potential for the RegistryCurator Agent. 

\subsection{RegistryCurator: Registry Evolution}
\label{subsec:curator}

The RegistryCurator Agent ensures Airavat's capabilities grow organically by identifying reusable patterns from successful workflows and proposing them for registry inclusion. Manual curation does not scale as the system generates more workflows, so the agent analyzes successful implementations to identify data processing utilities, analysis algorithms, and integration functions demonstrating utility beyond the original query context.

Critically, not all patterns merit generalization. The agent employs a validation-first strategy requiring proposed functions to pass four tests: (1) cross-scenario utility (value in 2-3 different use cases), (2) accuracy assessment (correct outputs on test cases), (3) integration compatibility (proper interaction with existing registry functions), and (4) edge case handling (appropriate behavior with invalid/boundary inputs). The agent must generate complete, executable validation code for every proposed function. This strict validation prevents registry bloat, which would increase token costs and degrade agent performance by overwhelming agents with excessive choices. Validation code runs manually to ensure quality control.

\section{Workflow Verification}
In this section, we detail the key components of Airavat that enable workflow verification.

\subsection{Knowledge Graph}

The Knowledge Graph condenses measurement research literature into a queryable resource for the Verification and Validation Engines. The construction pipeline addresses three challenges: preventing LLM hallucination during extraction, supporting both semantic search and relationship traversal, and processing thousands of papers cost-effectively.

\parab{Corpus Collection and Classification.} Airavat targets four ACM conferences  (SIGCOMM, SIGMETRICS, IMC, CoNEXT) and associated journals (POMACS, PACMNET) that capture foundational measurement methodologies and validation techniques from the past five decades. An automated scraper extracts DOIs and metadata (title, authors, page 
count, abstract) via Crossref API, with a 6-page threshold to exclude papers lacking sufficient methodological detail.
To focus on measurement research, a two-stage classification pipeline categorizes papers into 24 measurement research areas using a sentence transformer, with low-confidence papers undergoing secondary classification via a zero-shot classifier. This prioritizes efficiency—sentence transformers process the entire corpus while the expensive zero-shot classifier handles uncertain cases—achieving comparable accuracy to cloud-based LLMs at no cost with local LLMs.

\parab{Content Extraction.} 
Extracting structured information requires parsing PDFs and identifying content across five pre-defined categories: problem statements, methodologies, data sources, baseline comparisons, and validations (example in Appendix Figure~\ref{fig:kg_output},~\ref{fig:kg_output_2}). 
Airavat uses GROBID to parse PDFs into structured TEI format. Initial single-model extraction encountered hallucination and category confusion. A two-stage architecture addresses this: LLama-3.1-8B first classifies which categories are present in each section, then LLama-3.3-70B extracts content from tagged sections, 
reducing confusion by restricting extraction to appropriate sections.

\parab{Graph Construction and Representation.} With the structured information, Airavat constructs a Neo4j knowledge graph encoding measurement domain knowledge through semantic embeddings and typed relationships. The entity model includes ten entity types: Papers, Problems, ResearchGaps, Approaches, PipelineSteps, Algorithms, Metrics, Parameters, Datasets, and Validations. Entities connect through typed relationships (detailed in Appendix~\ref{sec:kg-details}).

\parab{Benefits and Transferability.} The graph representation enables semantic search, efficient multi-hop traversal, and pattern discovery, serving as the foundation for Verification and Validation Engines to query precedents and assess methodological soundness. The system is domain-agnostic—expertise emerges from the graph rather than from hardcoded logic, enabling extensibility by adding relevant papers. Additionally, our graph construction pipeline allows "incremental updates," enabling researchers to incorporate the latest conference proceedings with minimal manual effort.

\subsection{Verification Engine}

Airavat's Verification Engine assesses the generated workflows, identifies critical gaps, and produces modifications grounded in established techniques. Assessing workflow quality is challenging: unlike algorithm optimization with automated verifiers or performance-based system optimizations, measurement workflows span diverse domains with varying requirements and typically cannot be evaluated by execution due to long-term data collection requirements, large-scale infrastructure needs, and often a lack of well-defined success metrics.

The Verification Engine addresses these challenges with a three-stage pipeline. The \textit{Evaluator} performs a systematic assessment of the workflow using the Knowledge Graph. The \textit{Selector} determines the optimal verification strategy based on the Evaluator's scores, selecting whether a given workflow merits direct use, requires enhancement, or benefits from combining with complementary workflows. Finally, the \textit{Synthesizer} generates refined workflows: either by improving individual workflows in enhancement mode or by combining multiple workflows in hybrid mode. Note that the verification here assesses methodological alignment with prior work rather than guaranteeing the correctness of conclusions.

\parab{Evaluator: Multi-Dimensional Assessment.} 
The Evaluator assesses workflows across five dimensions: literature alignment, novelty, feasibility, simplicity, and robustness. These complementary dimensions serve dual purposes: computing overall quality scores for comparison and revealing specific improvement opportunities for synthesis. The evaluator employs a three-stage pipeline. 

The first stage involves structural validation. This stage validates JSON Schema compliance for intermediate outputs (e.g., the QueryMind output schema, Fig.~\ref{fig:querymind_schema} in the Appendix) confirms that all sub-problems are addressed, verifies that the proposed registry functions exist, and ensures that workflow complexity matches requirements. Workflows that fail structural validation are rejected.

The second stage focuses on literature-grounded scoring using the Knowledge Graph. Three complementary assessment approaches provide different perspectives: (a) Problem-centric assessment searches for similar problems via vector similarity, validating whether methods, steps, datasets, and targets align with approaches used for similar problems (contributes to literature alignment and robustness); (b) Approach-centric assessment searches for similar methods across literature without problem constraints, detecting transferable cross-domain patterns (contributes to literature alignment, novelty, feasibility, simplicity, and robustness); (c) Collective assessment analyzes patterns frequently appearing in literature but missing from all workflows, identifying systematic gaps and generating system warnings rather than dimension scores. This stage also validates feasibility through sub-dimension checks (scale handling, complexity, edge-case coverage, error handling), rejecting workflows that fail minimum thresholds [Appendix Table~\ref{tab:parameters}]. 

Finally, the Evaluator assigns the final score through adaptive weighting, which dynamically adjusts dimension priorities based on problem characteristics. Novel problems receive increased novelty weight, well-studied problems receive increased literature weight, and scale-intensive problems receive increased feasibility weight. Final scoring produces rankings with comprehensive justifications documenting dimension breakdowns, strengths/weaknesses, and workflow-specific limitations. System warnings capturing collective gaps appear separately from workflow-specific issues.

\parab{Selector: Verification Strategy Selection}. The Selector determines the optimal verification strategy based on Evaluator scores. Workflows exceeding excellence thresholds (Appendix Table~\ref{tab:parameters}) receive immediate approval. Workflows within the good range trigger an enhancement evaluation, which examines dimension weaknesses and workflow-specific limitations. When it identifies actionable issues, synthesis proceeds in ``enhancement'' mode. Workflows below the good range always trigger synthesis. When multiple proposals exist with structural diversity exceeding thresholds, the Selector performs complementarity analysis. If workflows propose fundamentally different approaches with complementary strengths, synthesis proceeds in ``hybrid'' mode to combine elements. If workflows are very similar, enhancement mode applies to the top-scoring workflow. The Selector prepares structured input for the Synthesizer, including problem context, workflows to improve or combine, complete evaluation insights, and improvement guidance.

\parab{Synthesizer: Generating Improved Workflows.} Synthesizer converts structured input into improved workflows using the top-performing model identified by the Selector. In enhancement mode, prompts include problem context, dimension weaknesses with computation transparency, knowledge graph hints with advisory markers, and domain considerations encouraging practical reasoning. In hybrid mode, prompts emphasize coherent combination guided by complementarity analysis, with explicit attention to common weaknesses. After generation, Synthesizer performs response parsing with schema validation, then generates change documentation via a separate call that compares the original and synthesized workflows. This separation ensures both synthesis quality and documentation comprehensiveness. Synthesized workflows maintain schema compatibility with WorkflowScout outputs, enabling seamless handoff to SolutionWeaver.

\subsection{Validation Engine}

Validation using alternative techniques is critical in measurement research to ensure that collected data accurately reflect reality and measurement tools produce reliable results. Airavat's Validation Engine generates executable validation code for network measurement solutions by discovering applicable validation approaches from research literature and adapting them to specific problems. The engine operates through a pipeline grouped into three functional components: InsightEngine, Strategizer, and CodeGenerator.

This decomposition enables three key capabilities. First, different validation tasks demand different computational approaches. Second, the pipeline gracefully adapts to varying literature coverage, emphasizing creative synthesis when few existing studies directly address a problem. Third, producing inspectable outputs at each stage allows researchers to understand how validation plans were derived rather than receiving unexplained recommendations. 

\parab{InsightEngine: Problem Analysis and Knowledge Discovery.}
InsightEngine performs three integrated functions to enable the synthesis of the validation strategies. First, in problem characterization, a local LLM (LLama-3.1-8B) semantically classifies problem characteristics (prediction, detection, temporal/spatial patterns, or causal relationships) without relying on keyword matching. This enables recognition of problem types across different terminologies and guides subsequent knowledge graph queries and validation-type selection. InsightEngine also identifies high-risk components flagged in WorkflowScout's analysis that require additional validation. Rather than executing potentially unsafe code, InsightEngine uses static analysis to extract implementation details from the generated workflow. Specifically, it parses the Python code into an Abstract Syntax Tree (AST) to identify key elements such as data sources, analytical methods, pipeline steps, and output structures. This extracted information then informs the queries sent to the Knowledge Graph to retrieve relevant validation strategies.

Second, to balance query specificity and coverage, InsightEngine uses multidimensional querying across seven dimensions: similar problems, methods, data sources, analysis pipelines, validation keywords, suitable ground-truth datasets, and domain-specific metrics. The local LLM first extracts technical terms and validation keywords from the problem description and implementation. Each query returns papers with their validation metadata (methodologies, ground truth sources, metrics, limitations), enabling the discovery of applicable approaches even when no single paper directly addresses the problem.

The third function assesses problem novelty. InsightEngine computes semantic similarity between the current problem and retrieved validation approaches using embedding models, capturing relationships that keyword matching would miss. High similarity indicates well-studied problems amenable to adapting proven methods. Low similarity signals novel problems requiring creative synthesis. For borderline cases, the LLM assesses whether semantically similar papers address comparable problems or differ in critical assumptions.

\parab{Strategizer: Filtering and Adaptation.}  
Strategizer employs an LLM to critically evaluate retrieved validation approaches, determine applicability, adapt them to the current problem, and select complementary strategies. This requires sophisticated reasoning because literature approaches rarely apply directly—papers may assume different data availability, constraints, or measurement capabilities. 

Strategizer receives extensive context from InsightEngine: problem characteristics, SolutionWeaver's implementation details, knowledge graph results with relevance scores, high-risk components, available tools and datasets from the Registry, and the novelty assessment. The synthesis process emphasizes critical filtering, identifying which approaches are actually applicable and explicitly documenting why others are unsuitable, preventing acceptance of semantically similar but practically inapplicable validations.

Three filtering rules guide this process. (i) Ground truth comparison is only recommended when ground truth demonstrably exists in the registry and matches the problem requirements. Recommending unavailable or mismatched ground truth wastes effort and provides no validation value. (ii) Alternative method validation is only suggested when the alternative method has proven reliability documented in the literature. Comparing against an unverified alternative provides no confidence gain. (iii) Validation strategies must provide complementary rather than redundant perspectives, examining different aspects such as end-to-end correctness, component-level accuracy, and internal consistency rather than repeatedly checking the same properties. 

For each recommended strategy, Strategizer specifies its applicability with supporting literature, adaptation from the original approach, feasibility given available data and tools, metrics with interpretation guidance, and how it complements other strategies. Validation strategies must trace to specific papers from the database and align with verified data availability rather than fabricating approaches or references. The grounding in knowledge graph results, combined with explicit feasibility checking, reduces hallucination by constraining generation through factual anchors. 

\parab{CodeGenerator: Producing Executable Validation Code.} CodeGenerator translates the validation plan into executable code using an LLM. Validation strategies typically span diverse types requiring different implementation approaches. Ground truth comparison, alternative method validation, consistency checking, component testing, and sample verification each have distinct implementation requirements. Templates would impose a rigid structure and fail to support creative strategies tailored to novel problems. The LLM observes SolutionWeaver's code to match its style, including documentation standards and coding patterns, while adapting to diverse validation approaches.

\begin{table*}[t]
\centering
\small
\caption{Case Study Summary}
\label{tab:case_studies}
\begin{tabular}{@{}lp{2.85cm}cp{4cm}p{5.3cm}@{}}
\toprule
\textbf{Case Study} & \textbf{Problem} & \textbf{LoC} & \textbf{Key Capability} & \textbf{Results} \\
\midrule
1: Cable Impact & SeaMeWe-5 failure & $\sim$850 & Expert Solution Replication & Matches the expert solution \\[0.15em]
2: Disaster & Earthquake/hurricane & $\sim$700 & Workflow Simplicity & Appropriate expert-level minimal solution \\[0.15em]
3: Cascading & Europe-Asia cables & $\sim$1600 & 9-function orchestration & Multi-layer integration framework \\[0.15em]
4: Prefix2Org & Prefix-to-org mapping & 1500$\rightarrow$1700 & Domain Transferability + & 0\%$\rightarrow$70-75\% (vague query); \\
 & & (w/ verification) & Verification capabilities & 0\%$\rightarrow$90.9\% tags (improved query) \\
\bottomrule
\end{tabular}
\end{table*} 

\section{Implementation}

The prompts instruct all agents to generate Python code. 

\parab{Workflow Generation}: Airavat employs Claude Opus 4.5 \cite{claude_opus_45} and Claude Sonnet 4.5~\cite{claude_sonnet_45} for the core agents in workflow generation. State-of-the-art cloud models provide the best available performance for the core functionality. The prompts are model-agnostic and do not rely on vendor-specific capabilities.  However, empirical evaluation shows that Claude variants perform most consistently (\S~\ref{subsec:model-compare}).

\parab{Workflow Verification}: The Knowledge Graph construction pipeline uses the mpnet-base-v2~\cite{allmpnetbasev2} sentence transformer for categorizing papers and BART-MNLI~\cite{bartmnli} as the zero-shot classifier. GROBID~\cite{GROBID} is used to parse the papers, followed by LLama-3.1-8B for tagging categories in each section, and LLama-3.3-70B for content extraction. The construction itself relies Neo4j~\cite{neo4j_platform}. The resulting graph contains 35,719 nodes across 2,021 papers, spanning over 65,000 relationships across eight entity types (detailed statistics in Appendix~\ref{kg_characteristic}). The Verification and Validation Engines rely on four cloud-based LLMs (Claude Opus 4.5~\cite{claude_opus_45}, Claude Sonnet 4.5~\cite{claude_sonnet_45}, Gemini-3-Pro~\cite{gemini_3_pro}, Gemini-3-Flash~\cite{ gemini_3_flash}) for generating multiple variants in LLM-assisted stages and BAAI BGE-M3~\cite{chen2025m3embeddingmultilingualitymultifunctionalitymultigranularity} for embeddings.

\parab{Cost:} Generation agents use Claude models (Sonnet 4.5 and Opus 4.5). Standard workflow generation costs \$0.80-\$1.70 per run (Agents 1-3), while comprehensive evaluation with 12 workflow variants, verification, and validation costs \$4.50-\$7.00. Total 
expenditure for all case studies is \$21.50, demonstrating cost-effectiveness 
for research-scale evaluation (detailed cost breakdown in Appendix~\ref{sec:cost-analysis}). Generation time for all workflows was under 10 minutes, significantly shorter than days/weeks required by human experts.

\section{Evaluation: Generation } 
\label{case_studies}

We demonstrate Airavat's workflow generation capabilities using four case studies spanning infrastructure resilience and IP allocation analysis. All evaluations use Claude Opus 4.5 unless mentioned otherwise. We organize the subsections based on demonstrated capabilities: expert solution replication, judicious tool selection, novel problem solving, and domain transferability.  Table~\ref{tab:case_studies} summarizes our results. We evaluate workflows along three axes: architectural coherence, correctness relative to expert baselines, and robustness to known data artifacts.

\subsection{Expert Solution Replication}
\label{sec:esr}
\parab{Capability Under Test:} Can Airavat derive workflows functionally equivalent to expert-designed solutions without domain-specific architectural guidance \textit{and} without the target solution in the Registry/Knowledge Graph?

\parab{Airavat Query:} "Identify the impact of SeaMeWe-5 cable failure at a country level" (Cable Impact Analysis Case Study)

\parab{Background:} Xaminer~\cite{xaminer} is an open-source cross-layer Internet resilience analysis framework that addresses such queries by aggregating metrics across the cable, IP, and AS layers.

\parab{Setup:} We provide Airavat with the standard registry, including core measurement functions from Nautilus~\cite{nautilus}, an open-source cross-layer submarine cable mapping framework, but \textit{deliberately excluding} Xaminer. This setup ensures the system relies on analytical reasoning rather than following the pre-existing solution. The query requires understanding cable dependencies, extracting affected IP addresses, performing geographic mapping, and aggregating country-level impacts—analysis that traditionally requires domain expertise and manual framework integration.

\parab{Workflow Comparison.} Both systems perform physical topology analysis (landing point identification) and traffic topology analysis (IP link processing). Xaminer employs embedding modules that pre-aggregate cross-layer metrics at country and AS-level abstractions. Airavat instead uses multi-source evidence fusion, assigning confidence scores based on source agreement (landing points, IP geolocation, AS registration) and separately tracking direct versus indirect landing station impacts.

\begin{figure}
    \centering
    \includegraphics[width=\linewidth]{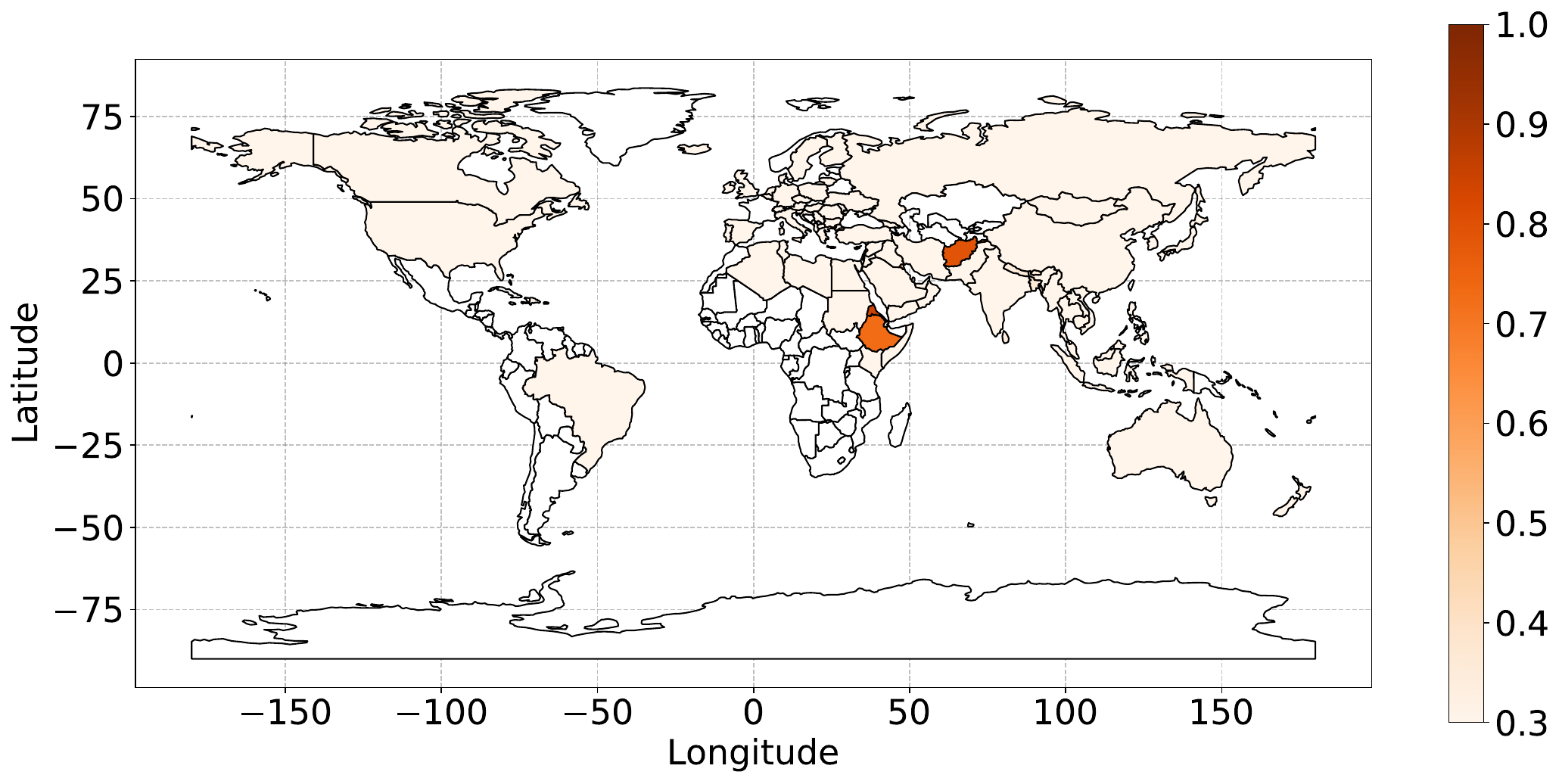}
    \caption{\small Country-level impact from SeaMeWe-5 cable failure. The heatmap shows the risk factor, computed as the fraction of IP addresses affected in a country relative to the total number of IP addresses mapped to that country. The results match the findings of Xaminer~\cite{xaminer}.}
    \label{fig:case_study1}
\end{figure}

\parab{Results.} Airavat successfully replicates Xaminer's analysis, producing perfectly matching results across all impact metrics (Figure~\ref{fig:case_study1}). Airavat generates 850 lines of Python code implementing the complete analysis pipeline using 4 registry functions (listed in Table~\ref{tab:registry_functions}). At the IP level, the most impacted countries are India (8,754 IPs), Malaysia (7,170), and Singapore (7,082). At the link level, the countries most affected by raw counts are Singapore (58.1K links), Germany (34.6K), and France (22.5K). At the AS level, the most impacted countries are Indonesia (869 ASes), Bangladesh (731), and Singapore (634). When examining normalized IP link impact (called risk factor by Xaminer), both systems identify Afghanistan, Eritrea, and Djibouti showing the highest normalized impact, followed by Nepal, Bhutan, and Bangladesh (Fig.~\ref{fig:case_study1}). The perfect agreement with Xaminer is expected: both systems ultimately invoke the same component functions and operate on identical input datasets. Additionally, the generated code includes comprehensive error handling, intermediate result validations, and clear documentation. This case study did not require any manual modifications.

\parab{Key Insight:} Airavat replicates expert-level analytical workflows through systematic reasoning about evidence fusion and confidence scoring, demonstrating that measurement expertise follows capturable compositional patterns.

\subsection{Judicious Tool Selection}
\label{sec:cs2}

\parab{Capability Under Test:} When multiple tools are available, will 
Airavat select minimal sufficient functionality or over-engineer 
solutions with unnecessary complexity?

\parab{Airavat Query:} "Identify the impact of severe earthquakes and hurricanes globally assuming a 10\% infrastructure failure probability" (Natural Disaster Analysis Case Study)

\parab{Setup:} We provide registry functions from multiple measurement frameworks, including Xaminer, to evaluate WorkflowScout's architectural decision-making. The challenge tests whether the system recognizes that Xaminer's event-processing capability can support multi-disaster analysis by systematically applying it to each disaster type, thereby avoiding unnecessary cross-framework orchestration.

\parab{Workflow Comparison.} Airavat demonstrates appropriate architectural restraint. Rather than orchestrating multiple specialized frameworks, the system identifies that Xaminer's event-processing function can handle both disaster types when applied systematically. The workflow processes earthquakes (MMI $\geq$ 3) and hurricanes (wind speed $\geq$ 75 knots) separately with 10\% failure probabilities, then merges results through union-based aggregation. While Xaminer makes a single call to generate combined results for both disasters, Airavat's separation enables disaster-specific impact statistics while maintaining architectural simplicity. Importantly, while the Xaminer implementation uses 7 registry functions (Table~\ref{tab:registry_functions}) to establish the analytical foundation, Airavat recognizes that one function can handle both types of disaster without requiring separate frameworks for each scenario.

\begin{figure}
    \centering
    \includegraphics[width=\linewidth]{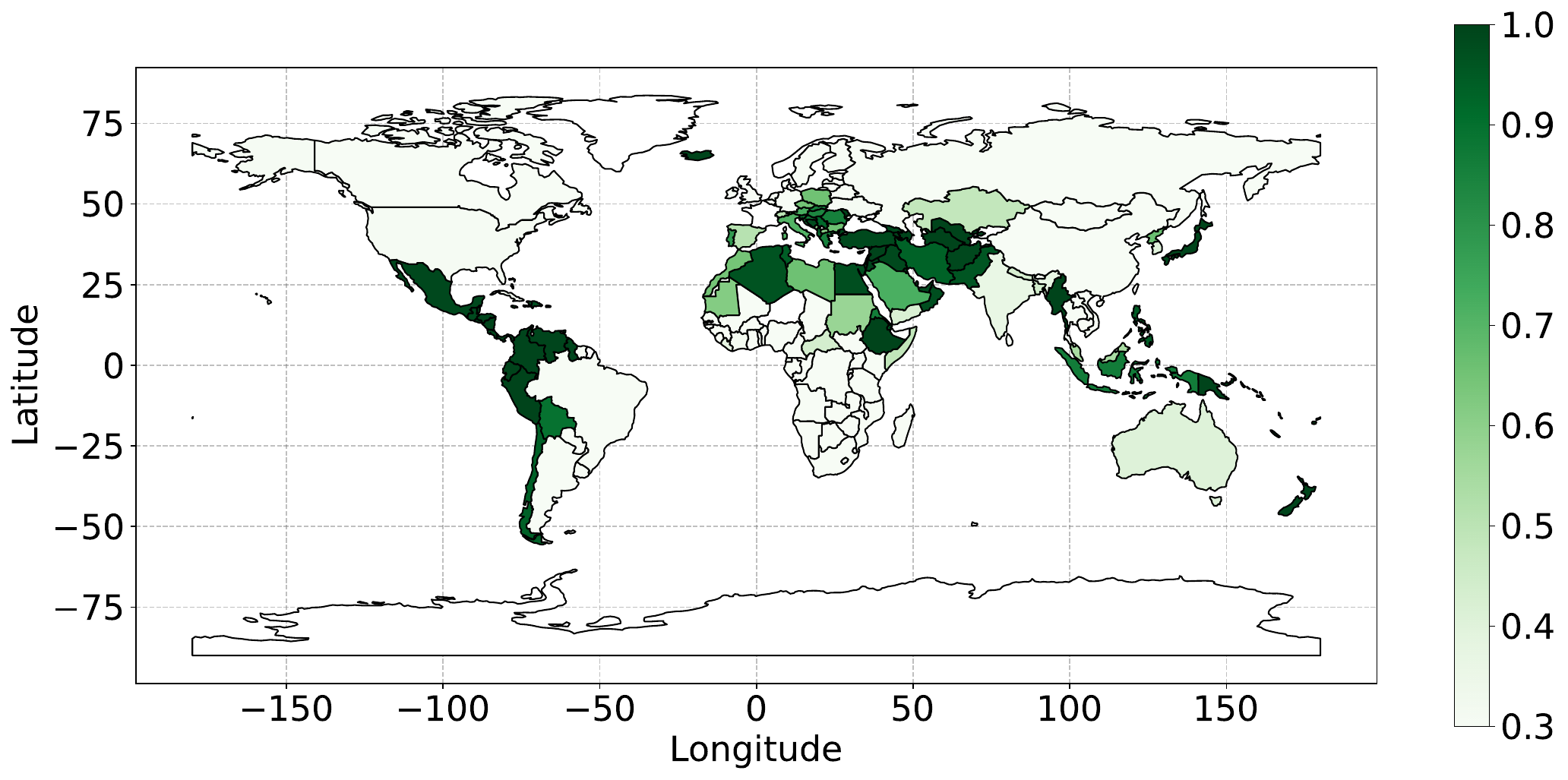}
    \caption{\small Global cable infrastructure impact from earthquakes and hurricanes. The heatmap shows the risk factor, computed as the fraction of IP links affected in a country relative to the total number of IP links in that country. The results match the findings of Xaminer~\cite{xaminer}.}
    \label{fig:case_study2}
\end{figure}

\parab{Results.} Airavat generates approximately 700 lines of Python code to implement multi-disaster analysis. Execution against real disaster data produces results perfectly matching Xaminer's multi-disaster impact analysis (Figure~\ref{fig:case_study2}).

\parab{Key Insight:} Airavat demonstrates appropriate architectural judgment by avoiding unnecessary complexity and selecting solutions that meet requirements without over-engineering.

\subsection{Novel Problem Solving}
\label{subsec:cascading}

\parab{Capability Under Test:} Can Airavat tackle research problems 
without established solutions, enabling analyses previously impractical due to integration complexity?

\parab{Airavat Query:} "Analyze the cascading effects of submarine cable failures between Europe and Asia" (Cascading Failure Analysis Case Study)

\parab{Challenge:} Unlike previous case studies with expert-designed solutions for validation, cascading failure analysis across continents has no established implementation. Manual development would require:  (i) expertise across infrastructure mapping, impact analysis, and AS  dependency tracking frameworks; (ii) days of integration engineering;  and (iii) specialized knowledge of cross-layer synthesis techniques. This barrier makes such analysis impractical for most researchers.

\parab{Setup:} We evaluate whether Airavat can orchestrate complex multi-framework workflows for exploratory research. The analysis requires integration across infrastructure mapping, AS-level dependency tracking, and cross-layer synthesis—traditionally requiring substantial manual engineering.

\parab{Workflow Structure.} Airavat produces a 1,600-line implementation orchestrating 9 registry functions across three analytical layers (Table~\ref{tab:registry_functions}):

\noindent
\textit{Physical Infrastructure Analysis}. Identifies submarine cables connecting Europe and Asia through geographic filtering of landing points. Using country-to-cable graph data and landing point mappings, it employs a MERGE strategy combining geographic filtering with cable-name-based identification.

\noindent
\textit{AS-Level Dependency Analysis.} Captures cascading effects through autonomous system relationships. Extracts AS numbers from affected cables, loads AS dependency graphs, and implements graph traversal to trace secondary impacts, distinguishing primary from 
secondary failures.

\noindent
\textit{Country-Level Impact Assessment.} Integrating physical 
and logical layers, it consolidates cable segments with IP-to-ASN mappings and geolocation data, creates an indexed 
embeddings for country/AS queries, and simulates multiple failures 
scenarios producing quantitative impacts across cable segments, IP 
links, IPs, AS links, and ASes.

\parab{Results.} Without ground truth, we assess workflow quality through architectural coherence, tool integration appropriateness, and analytical reasoning. The generated solution demonstrates: (1) correct understanding of measurement tool capabilities across frameworks, (2) appropriate cross-layer integration, and (3) specialized domain reasoning for cascade analysis. Execution against real infrastructure and routing data produces comprehensive vulnerability assessments that provide actionable starting points for further investigation. Figures~\ref{fig:cascade_workflow1},~\ref{fig:cascade_workflow2},~\ref{fig:cascade_workflow3}, and~\ref{fig:cascade_workflow4} visually show the workflow layers generated by Airavat.

\parab{Key Insight:} Airavat handles complex multi-framework integration for never-solved-before problems, lowering barriers to sophisticated exploratory analysis while maintaining research-quality reasoning across network layers.

\subsection{Domain Transferability}
\label{subsec:transfer}
\parab{Capability Under Test:} Can Airavat adapt to different measurement problems?

\parab{Airavat Query:} "Map BGP-routed prefixes to owner organizations, identifying Direct-Owner allocations, Delegated Customer chains, and organizational consolidation for address blocks scattered across WHOIS records" (Prefix-to-Organization Mapping Case Study)

\parab{Background:} Prefix2Org~\cite{prefix2org} maps BGP prefixes to organizations, distinguishing Direct Owners (provider-independent allocations) from Delegated Customers (sub-delegations) through WHOIS parsing, longest-prefix-match using radix tries, and organizational clustering across heterogeneous WHOIS formats from all RIRs.

\parab{Setup and Challenge:} This case study focuses on IP allocation analysis (a significantly different measurement problem from prior case studies on infrastructure resilience) under deliberately more stringent constraints. Registry functions provide raw BGP/WHOIS data with minimal preprocessing (Table~\ref{tab:registry_functions}), unlike previous case studies where substantial processing occurs upstream. We do not define key terms in the query, such as "Direct Owner" and "Delegated Customer". This tests whether LLM reasoning can infer semantic distinctions from context alone. We generate 12 workflow variants using four models (Claude Sonnet 4.5, Claude Opus 4.5, Gemini-3-Pro, Gemini-3-Flash) at three temperature settings (0.0, 0.5, 1.0) and evaluate against Prefix2Org monthly ground truth data covering over 1 million BGP-routed prefixes.

\parab{Domain Transfer Results.} WorkflowScout successfully captures high-level workflow methodology for this new domain, generating specifications that incorporate domain-appropriate techniques including radix trie construction for longest-prefix-match operations, WHOIS format parsing across heterogeneous RIR schemas, and organizational name matching for entity consolidation. The generated workflows process completely different data sources (bulk WHOIS records, BGP dumps, RPKI certificates) using methodology distinct from cable-based analysis, demonstrating that Airavat adapts to new measurement domains rather than simply templating from previous solutions.

All 12 generated workflows execute successfully and produce output in expected formats, validating domain transfer at the architectural and implementation levels.  However, comparison against Prefix2Org ground truth reveals 0\% correct mappings. Every workflow contains an identical critical bug: failing to filter 0.0.0.0/0 default routes from BGP dumps before hierarchical tree traversal operations, causing all prefixes to incorrectly map to catch-all allocations. 
This demonstrates that subtle domain-specific 
data quality requirements that exist in tacit expert knowledge rather than formal 
specifications require verification mechanisms to detect and address them. We discuss how the Verification Engine helps improve the result next.

\section{Evaluation: Verification}
\label{sec:verify-results}

Using the Prefix2Org case study~\cite{prefix2org}, which demonstrated that all generated workflows failed despite syntactically correct code (\S~\ref{subsec:transfer}), we demonstrate the verification capabilities of Airavat. Specifically, we illustrate the following capabilities: bug detection, comparative model evaluation, including the filtering of faulty workflows, and the impact of query clarity on performance. Verification success is measured by the detection of known methodological flaws and improvement in downstream accuracy.

\subsection{Literature-Grounded Bug Detection}
\label{subsec:bug}

\parab{Goal:} Identify subtle methodological flaws that execution testing cannot detect—bugs that produce syntactically correct code executing successfully yet yielding incorrect measurement results.

\parab{Experimental Setup:} We generate workflows using four models (Claude Sonnet 4.5, Claude Opus 4.5, Gemini-3-Pro, Gemini-3-Flash), at three temperatures (0.0, 0.5, 1.0), producing 12 configurations. Each configuration is run once, yielding a total of 12 generated workflows.

\parab{The Bug:} Without verification, all 12 baseline workflows contained an identical critical bug: failing to filter 0.0.0.0/0 default routes before hierarchical tree traversal operations, causing every prefix to match the catch-all route. Additionally, workflows failed to filter overly broad WHOIS allocations (IPv4 < /8, IPv6 < /16) representing RIR administrative  records rather than operational allocations. This preprocessing requirement exists in tacit expert knowledge, but rarely appears in published algorithmic descriptions—making it invisible to LLMs despite correct high-level methodology.

\parab{Bug Detection:} The Verification Engine detects and repairs bugs in three stages. The evaluator scored workflows against the quality dimensions, assigning scores below the excellence threshold, triggering enhancement mode. It then compared the workflows against best practices embedded in the Knowledge Graph, identified systematic data quality issues, and generated two warnings. The first warning concerns BGP data quality and the need to remove unwanted prefixes, which correctly identifies the bug under consideration. The second is about inferred topology validation, which proved less relevant. These warnings are derived from literature patterns: papers in the knowledge graph discuss default route filtering, establishing it as a domain best practice. Synthesizer uses these warnings to generate enhanced workflows.

\parab{Results:} Workflows improved from 0\% accuracy to operational correctness with one targeted fix guided by automated detection. Synthesizer correctly fixed the 0.0.0.0/0 bug for BGP prefixes, but issued a warning for the same issue for WHOIS records. The synthesized workflow also automatically implemented bogon prefix filtering and invalid AS record elimination, with additional warnings flagging suspicious patterns (e.g., large numbers of ASNs assigned to the U.S. DoD). Due to a warning for WHOIS, unlike an updated workflow in the case of BGP, LLM-guided manual correction of the WHOIS 0.0.0.0/0 record was required. Note that this correction is also fully automated without manual intervention when the query has better clarity (\S~\ref{sec:clarity}).

\parab{Key Insight.} Literature-grounded verification detects bugs missed by standard execution-based testing by comparing generated workflows against best practices extracted from published research. When full automatic repair isn't possible, the system provides graceful degradation: explicit warnings with context enable quick manual correction rather than leaving bugs undetected. 

\subsection{Comparative Model Evaluation}
\label{subsec:model-compare}

\parab{Goal:} Systematically evaluate model-specific failure patterns in the Verification Engine.

\parab{Experimental Design:} We consider the same 12 configurations in \S~\ref{subsec:bug} with three runs per configuration generating a total of 36 workflows.

\parab{The Bug:} We consider the same bug as \S~\ref{subsec:bug}.

\parab{Model-Specific Failure Detection:} The Verification Engine's three-stage pipeline detected systematic model-specific patterns. Structural validation automatically filtered either Gemini-3-Pro or Gemini-3-Flash at a temperature of 0.0 in all three runs due to malformed outputs (incomplete schema compliance, missing fields, invalid JSON).  This aligns with Google's warning against very low temperatures for Gemini-3 models~\cite{google_gemini3_temperature}. Dimension scoring revealed robustness failures across the remaining Gemini configurations. Most were tagged infeasible due to missing edge case handling and error scenarios documented in prior IP allocation research. This matches Google's documentation indicating Gemini's preference for direct answers over comprehensive robustness~\cite{google_gemini3_prompt}. In 1 of 3 runs, Opus 4.5 (temperature 1.0) was tagged infeasible for excessive complexity exceeding literature norms. Quality ranking consistently placed Sonnet 4.5 variants highest across all runs (scores 68-72), validating our design choice to employ Claude variants for the agent pipeline.

\parab{Key Insight:} The Verification Engine identifies model-specific failure patterns through empirical assessment rather than requiring manual expertise about LLM-specific behaviors and limitations. Systematic evaluation across architectures and temperatures reveals that model selection significantly impacts workflow quality, with structural validation and robustness scoring providing objective selection criteria.

\subsection{Query Clarity Impact Assessment}
\label{sec:clarity}

\parab{Goal:} Evaluate how query precision affects LLM reasoning quality during verification by comparing workflows generated from vague versus refined queries.

\parab{Experimental Design:} We compare two query variants for the same Prefix2Org task. \textit{(i) Vague query (baseline):} Airavat Query in \S~\ref{subsec:transfer}. This query deliberately omits definitions for "Direct Owner" and 
"Delegated Customer". \textit{(ii) Refined query:} We add targeted hints without complete definitions for Direct Owners and Delegated Customers. 

\parab{Performance Impact of Query Clarity:} With the vague query (no Direct/Delegated definitions), the corrected workflow generated by the Verification Engine achieved 97.6\% overlap for origin AS identification and 60-65\% for Direct Owner/Delegated Customer (DO/DC) WHOIS tag classification, improving to 70-75\% after excluding incomplete LACNIC data. This represents substantial improvement from 0\% correctness without verification. 

With the refined query before verification, all 12 workflows still contained the identical 0.0.0.0/0 bug. After verification, the overlap for origin AS identification remained the same, and the DO/DC WHOIS tag identification improved to 90.9\% (correctly identified 20 of 22 tags, with only two LACNIC tags misclassified). The workflow also included RPKI validation as a fallback (similar to the paper~\cite{prefix2org}) for low-confidence tags. For this improved query, both the BGP and WHOIS 0.0.0.0/0 issue was automatically detected and fixed during synthesis, requiring no manual corrections. 

\parab{Key Insight:} Query precision dramatically impacts LLM semantic 
reasoning quality for problems requiring domain-specific interpretation beyond technical implementation. Even partial clarification (hints rather than complete definitions) improves ownership classification accuracy from  60-65\% to 90.9\%, demonstrating that query engineering complements verification in achieving high-quality workflows.

\section{Evaluation: Validation}
\label{sec:validate-results}

We demonstrate the Validation Engine capabilities with the Cable Impact Analysis Case Study (\S~\ref{sec:esr}). The workflow analyzes country-level impact from SeaMeWe-5 cable failure, producing metrics spanning cable segments, IP links, ASes, and geographic distribution. For this evaluation, we generate a new knowledge graph by removing Xaminer paper~\cite{xaminer} to prevent the validation engine from directly identifying techniques from the paper.

The Validation Engine generated nine validation strategies spanning multiple validation types: three system-level validations, three component validations, one consistency validation, and two synthesized validations. We show how the engine identifies validation strategies in the existing literature and develops new ones when precedents are absent.

\parab{[V1] System-Level Validation:  Landing Point Country Coverage.} InsightEngine's problem-centric knowledge graph queries identified POPsicle's~\cite{popsicle} validation approach comparing network topology predictions against authoritative sources. The Strategizer adapted this to submarine cable validation, proposing comparison of the workflow's country list against publicly documented SeaMeWe-5 landing points from TeleGeography and SubmarineCableMap~\cite{submarine_map}.

\parab{[V2] Component Validation: Cable Segment Identification.} InsightEngine flagged the Cable Segment Identification component as high-risk based on WorkflowScout's analysis, triggering targeted component-level validation. The Strategizer adapted SyslogDigest's template validation approach~\cite{syslogdigest} (comparing identified patterns against ground truth) to cable identification, proposing dual verification through string matching across naming variants (SeaMeWe-5, SMW5, SMW-5) and landing point pair matching. The strategy validates that identified segments connect known landing point pairs and form geographically coherent routes, isolating cable identification accuracy from downstream aggregation errors.

\parab{[V3] Consistency Validation:  Multi-Source Impact Metric Consistency. } InsightEngine's approach-centric queries identified Tessellation's multi-source traffic attribution validation~\cite{tessellation}. The Strategizer adapted this to country attribution consistency checking across three independent sources: IP geolocation, ASN mappings, and landing point data. The strategy computes consistency scores (proportion of sources agreeing) and identifies countries with high versus low source agreement, providing confidence indicators for impact assessments when single ground truth is unavailable.

\parab{[V4] Synthesized Validation:  Historical Cable Outage Comparison. } InsightEngine's novelty assessment indicated no direct precedent for submarine cable failure impact validation, triggering creative synthesis. The Strategizer synthesized this approach by adapting temporal validation patterns from network disruption detection literature, proposing application of the workflow to documented historical outages (2008 Mediterranean cuts, 2020 AAE-1 issues) and comparison of predicted impacts against reported impacts. Critically, the strategy leverages IODA traffic measurements and RIPE Atlas data during historical cable failures—matching the validation methodology employed in the original Xaminer paper~\cite{xaminer} for infrastructure resilience analysis. This demonstrates the Validation Engine's ability to discover and propose validation approaches that align with established measurement research practices.

\parab{Methodological Quality Assessment.} The generated strategies exhibit multiple quality indicators. First, literature grounding ensures strategies adapt proven validation patterns rather than fabricating approaches. Second, complementarity ensures strategies examine different correctness dimensions—V1 validates geographic scope, V2 validates data integration quality, V3 validates foundational component accuracy, V4 provides external validation through historical comparison. Third, feasibility assessment ensures all strategies use data available in the registry or publicly accessible sources. Fourth, explicit adaptation justifications document how literature approaches transfer to the specific problem context. 

\section{Discussion and Related work}

\parab{Generalization.} We developed agent prompts through iterative refinement, encoding domain-agnostic reasoning patterns (e.g., problem decomposition, constraint evaluation) while deliberately excluding measurement-specific heuristics. Domain expertise resides in the Knowledge Graph and Registry rather than agent logic, enabling transferability through knowledge graph population. 

Within the Internet measurement domain, Airavat's architecture generalizes across sub-domains, as demonstrated by our case study (\S~\ref{subsec:transfer}). We expect Airavat's capabilities to readily generalize across a broader range of sub-problems, including network security analysis and performance debugging. The knowledge graph and agent reasoning patterns transfer directly to these domains by ingesting relevant research papers and domain-specific tools. Beyond Internet measurement, generalization to fundamentally different domains requires domain-specific knowledge graph construction and registry curation. Additionally, measurement-specific components, such as the five complexity dimensions used for query decomposition, may require adaptation for other fields.

\parab{Emerging Standards and Scalability.} The emergence of agent communication protocols such as Model Context Protocol (MCP)~\cite{mcp_anthropic} and Agent-to-Agent protocol (A2A)~\cite{a2a_protocol}  presents opportunities for standardizing AI agent interactions with measurement tools. MCP's server-client design could provide unified interfaces for tool interaction, dramatically simplifying registry maintenance through automatic capability discovery and standardized interaction patterns. A2A protocols could formalize communication between Airavat's specialized agents, enabling more robust task delegation and state management. However, realizing these benefits requires widespread protocol adoption across both the measurement tool and AI agent ecosystems. A related challenge is maintaining registry 
accuracy as tools evolve. Future work could employ specialized LLM agents to 
automatically analyze codebases and monitor tool repositories, reducing manual 
maintenance overhead while improving scalability.

\parab{Limitations.} Despite demonstrated capabilities, Airavat exhibits fundamental limitations. While it can be fully automated for well-known problems, it can only serve as a co-pilot for never-seen-before problems. The Registry requires manual curation; while the RegistryCurator Agent identifies reusable patterns, all additions require expert validation to maintain quality standards. While the knowledge graph captures a broad slice of measurement literature, it may miss informal best practices. The verification engine can only detect methodological flaws documented in existing literature. The system cannot generate validation strategies for workflows requiring long-term longitudinal studies, large-scale infrastructure deployment, or proprietary datasets unavailable in public repositories. Finally, query precision can dramatically impact accuracy, as demonstrated in \S~\ref{sec:clarity}, requiring both experts and non-experts to be precise in their interactions.

\parab{Multi-agent LLM systems} decompose 
complex problems into specialized subtasks. Prominent frameworks include  AutoGen~\cite{wu2023autogenenablingnextgenllm}, MetaGPT~\cite{hong2024metagptmetaprogrammingmultiagent}, LangGraph~\cite{langgraph2026}, and CrewAI~\cite{crewai2026}. However, multi-agent  systems exhibit notable pitfalls including role violations, input conflicts, and  incomplete verification~\cite{cemri2025multiagentllmsystemsfail}. Airavat addresses 
these challenges through systematic quality assurance mechanisms. 

\parab{\textbf{Agentic Workflows in Networked Systems Research.}} Recent work applies LLMs and agentic AI to networked systems: NetLLM~\cite{wu2024netllm} for networking tasks, NetConfEval~\cite{wang2024netconfeval} for configuration, Confucius~\cite{wang2025confucius}  for network management, and system optimization~\cite{adrs,glia,ro2025sherlockreliableefficientagentic}.  However, none address measurement research's unique challenges of  workflow  generation with systematic verification, which Airavat provides.

\parab{\textbf{AI for Science.}} AI's role in scientific discovery spans multiple fields 
and autonomy levels~\cite{zheng2025automationautonomysurvey,lu2024aiscientistfullyautomated, 
yamada2025aiscientistv2workshoplevelautomated, gottweis2025aicoscientist}, from passive 
assistance to fully autonomous research. Achieving autonomous scientific discovery requires 
advances across problem identification, hypothesis formulation, experiment design, execution, 
analysis, and iterative refinement. Our work contributes by developing agentic workflows 
for network measurement research that demonstrate how multi-LLM collaboration can address 
domain-specific scientific challenges while maintaining interpretability and human oversight. 

\section{Conclusion}

Internet measurement research requires sophisticated tool integration and rigorous validation. Airavat demonstrates how agentic AI systems can automate both workflow generation and literature-grounded verification through specialized agents and engines operating on a knowledge graph. Our evaluation shows that agentic systems generate expert-level workflows and identify methodological flaws missed by standard execution-based testing. We view Airavat as a force multiplier for experts and a scaffolding tool for non-specialists. Looking forward, we envision Airavat enabling a new mode of measurement research—where hypotheses, workflows, verification strategies, and validation plans co-evolve interactively—lowering the barrier to rigorous Internet measurement while preserving the methodological discipline built over decades of community effort.

\bibliographystyle{ACM-Reference-Format}
\bibliography{reference}

@misc{zheng2025automationautonomysurvey,
      title={From Automation to Autonomy: A Survey on Large Language Models in Scientific Discovery}, 
      author={Tianshi Zheng and Zheye Deng and Hong Ting Tsang and Weiqi Wang and Jiaxin Bai and Zihao Wang and Yangqiu Song},
      year={2025},
      eprint={2505.13259},
      archivePrefix={arXiv},
      primaryClass={cs.CL},
      url={https://arxiv.org/abs/2505.13259}, 
}

@misc{lu2024aiscientistfullyautomated,
      title={The AI Scientist: Towards Fully Automated Open-Ended Scientific Discovery}, 
      author={Chris Lu and Cong Lu and Robert Tjarko Lange and Jakob Foerster and Jeff Clune and David Ha},
      year={2024},
      eprint={2408.06292},
      archivePrefix={arXiv},
      primaryClass={cs.AI},
      url={https://arxiv.org/abs/2408.06292}, 
}

@misc{yamada2025aiscientistv2workshoplevelautomated,
      title={The AI Scientist-v2: Workshop-Level Automated Scientific Discovery via Agentic Tree Search}, 
      author={Yutaro Yamada and Robert Tjarko Lange and Cong Lu and Shengran Hu and Chris Lu and Jakob Foerster and Jeff Clune and David Ha},
      year={2025},
      eprint={2504.08066},
      archivePrefix={arXiv},
      primaryClass={cs.AI},
      url={https://arxiv.org/abs/2504.08066}, 
}

@misc{gottweis2025aicoscientist,
      title={Towards an AI co-scientist}, 
      author={Juraj Gottweis and Wei-Hung Weng and Alexander Daryin and Tao Tu and Anil Palepu and Petar Sirkovic and Artiom Myaskovsky and Felix Weissenberger and Keran Rong and Ryutaro Tanno and Khaled Saab and Dan Popovici and Jacob Blum and Fan Zhang and Katherine Chou and Avinatan Hassidim and Burak Gokturk and Amin Vahdat and Pushmeet Kohli and Yossi Matias and Andrew Carroll and Kavita Kulkarni and Nenad Tomasev and Yuan Guan and Vikram Dhillon and Eeshit Dhaval Vaishnav and Byron Lee and Tiago R D Costa and José R Penadés and Gary Peltz and Yunhan Xu and Annalisa Pawlosky and Alan Karthikesalingam and Vivek Natarajan},
      year={2025},
      eprint={2502.18864},
      archivePrefix={arXiv},
      primaryClass={cs.AI},
      url={https://arxiv.org/abs/2502.18864}, 
}

@misc{google_gemini3_temperature,
  title        = {Gemini 3 Developer Guide},
  author       = {{Google}},
  year         = {2025},
  howpublished = {\url{https://ai.google.dev/gemini-api/docs/gemini-3##temperature}},
}

@misc{google_gemini3_prompt,
  title        = {Gemini 3 Developer Guide},
  author       = {{Google}},
  year         = {2025},
  howpublished = {\url{https://ai.google.dev/gemini-api/docs/gemini-3##prompting_best_practices}},
}

@article{nautilus,
author = {Ramanathan, Alagappan and Abdu Jyothi, Sangeetha},
title = {Nautilus: A Framework for Cross-Layer Cartography of Submarine Cables and IP Links},
year = {2023},
issue_date = {December 2023},
publisher = {Association for Computing Machinery},
address = {New York, NY, USA},
volume = {7},
number = {3},
url = {https://doi.org/10.1145/3626777},
doi = {10.1145/3626777},
journal = {Proc. ACM Meas. Anal. Comput. Syst.},
month = dec,
articleno = {46},
numpages = {34}
}

@article{xaminer,
author = {Ramanathan, Alagappan and Sankaran, Rishika and Abdu Jyothi, Sangeetha},
title = {Xaminer: An Internet Cross-Layer Resilience Analysis Tool},
year = {2024},
issue_date = {March 2024},
publisher = {Association for Computing Machinery},
address = {New York, NY, USA},
volume = {8},
number = {1},
url = {https://doi.org/10.1145/3639042},
doi = {10.1145/3639042},
journal = {Proc. ACM Meas. Anal. Comput. Syst.},
month = feb,
articleno = {16},
numpages = {37}
}

@misc{multiple_providers_aae1_outage,
author = {},
year={2022},
title = {{AAE-1 cable cut causes widespread outages in Europe, East Africa, Middle East, and South Asia - DCD}},
howpublished = {\url{https://www.datacenterdynamics.com/en/news/aae-1-cable-cut-causes-widespread-outages-in-europe-east-africa-middle-east-and-south-asia/}},
}

@inproceedings{wu2024netllm,
author = {Wu, Duo and Wang, Xianda and Qiao, Yaqi and Wang, Zhi and Jiang, Junchen and Cui, Shuguang and Wang, Fangxin},
title = {NetLLM: Adapting Large Language Models for Networking},
year = {2024},
isbn = {9798400706141},
publisher = {Association for Computing Machinery},
address = {New York, NY, USA},
url = {https://doi.org/10.1145/3651890.3672268},
doi = {10.1145/3651890.3672268},
booktitle = {Proceedings of the ACM SIGCOMM 2024 Conference},
pages = {661–678},
numpages = {18},
location = {Sydney, NSW, Australia},
series = {ACM SIGCOMM '24}
}

@article{wang2024netconfeval,
author = {Wang, Changjie and Scazzariello, Mariano and Farshin, Alireza and Ferlin, Simone and Kosti\'{c}, Dejan and Chiesa, Marco},
title = {NetConfEval: Can LLMs Facilitate Network Configuration?},
year = {2024},
issue_date = {June 2024},
publisher = {Association for Computing Machinery},
address = {New York, NY, USA},
volume = {2},
number = {CoNEXT2},
url = {https://doi.org/10.1145/3656296},
doi = {10.1145/3656296},
journal = {Proc. ACM Netw.},
month = jun,
articleno = {7},
numpages = {25}
}

@inproceedings{wang2025confucius,
author = {Wang, Zhaodong and Lin, Samuel and Yan, Guanqing and Ghorbani, Soudeh and Yu, Minlan and Zhou, Jiawei and Hu, Nathan and Baruah, Lopa and Peters, Sam and Kamath, Srikanth and Yang, Jerry and Zhang, Ying},
title = {Intent-Driven Network Management with Multi-Agent LLMs: The Confucius Framework},
year = {2025},
isbn = {9798400715242},
publisher = {Association for Computing Machinery},
address = {New York, NY, USA},
url = {https://doi.org/10.1145/3718958.3750537},
doi = {10.1145/3718958.3750537},
booktitle = {Proceedings of the ACM SIGCOMM 2025 Conference},
pages = {347–362},
numpages = {16},
location = {S\~{a}o Francisco Convent, Coimbra, Portugal},
series = {SIGCOMM '25}
}

@misc{wu2023autogenenablingnextgenllm,
      title={AutoGen: Enabling Next-Gen LLM Applications via Multi-Agent Conversation}, 
      author={Qingyun Wu and Gagan Bansal and Jieyu Zhang and Yiran Wu and Beibin Li and Erkang Zhu and Li Jiang and Xiaoyun Zhang and Shaokun Zhang and Jiale Liu and Ahmed Hassan Awadallah and Ryen W White and Doug Burger and Chi Wang},
      year={2023},
      eprint={2308.08155},
      archivePrefix={arXiv},
      primaryClass={cs.AI},
      url={https://arxiv.org/abs/2308.08155}, 
}

@misc{hong2024metagptmetaprogrammingmultiagent,
      title={MetaGPT: Meta Programming for A Multi-Agent Collaborative Framework}, 
      author={Sirui Hong and Mingchen Zhuge and Jiaqi Chen and Xiawu Zheng and Yuheng Cheng and Ceyao Zhang and Jinlin Wang and Zili Wang and Steven Ka Shing Yau and Zijuan Lin and Liyang Zhou and Chenyu Ran and Lingfeng Xiao and Chenglin Wu and Jürgen Schmidhuber},
      year={2024},
      eprint={2308.00352},
      archivePrefix={arXiv},
      primaryClass={cs.AI},
      url={https://arxiv.org/abs/2308.00352}, 
}

@misc{langgraph2026,
  title        = {LangGraph: Agent Orchestration Framework},
  howpublished = {\url{https://www.langchain.com/langgraph}},
  year         = {2026},
  organization = {LangChain, Inc.}
}

@misc{crewai2026,
  title        = {CrewAI},
  howpublished = {\url{https://www.crewai.com/}},
  year         = {2026},
  organization = {CrewAI, Inc.}
}

@misc{cemri2025multiagentllmsystemsfail,
      title={Why Do Multi-Agent LLM Systems Fail?}, 
      author={Mert Cemri and Melissa Z. Pan and Shuyi Yang and Lakshya A. Agrawal and Bhavya Chopra and Rishabh Tiwari and Kurt Keutzer and Aditya Parameswaran and Dan Klein and Kannan Ramchandran and Matei Zaharia and Joseph E. Gonzalez and Ion Stoica},
      year={2025},
      eprint={2503.13657},
      archivePrefix={arXiv},
      primaryClass={cs.AI},
      url={https://arxiv.org/abs/2503.13657}, 
}

@misc{ro2025sherlockreliableefficientagentic,
      title={Sherlock: Reliable and Efficient Agentic Workflow Execution}, 
      author={Yeonju Ro and Haoran Qiu and Íñigo Goiri and Rodrigo Fonseca and Ricardo Bianchini and Aditya Akella and Zhangyang Wang and Mattan Erez and Esha Choukse},
      year={2025},
      eprint={2511.00330},
      archivePrefix={arXiv},
      primaryClass={cs.MA},
      url={https://arxiv.org/abs/2511.00330}, 
}

@misc{allmpnetbasev2,
  author = {Reimers, Nils and Gurevych, Iryna},
  title = {all-mpnet-base-v2},
  year = {2021},
  howpublished = {\url{https://huggingface.co/sentence-transformers/all-mpnet-base-v2}},
}

@misc{bartmnli,
  author = {Facebook AI},
  title = {bart-large-mnli},
  year = {2020},
  howpublished = {\url{https://huggingface.co/facebook/bart-large-mnli}},
}

@misc{claude_opus_45,
  title        = {Claude Opus 4.5},
  author       = {{Anthropic}},
  year         = {2025},
  howpublished = {\url{https://www.anthropic.com/news/claude-opus-4-5}}
}

@misc{claude_sonnet_45,
  title        = {Claude Sonnet 4.5},
  author       = {{Anthropic}},
  year         = {2025},
  howpublished = {\url{https://www.anthropic.com/news/claude-sonnet-4-5}}
}

@misc{gemini_3_pro,
  title        = {Gemini 3 Pro},
  author       = {{Google DeepMind}},
  year         = {2025},
  howpublished = {\url{https://deepmind.google/models/gemini/pro/}}
}

@misc{gemini_3_flash,
  title        = {Gemini 3 Flash},
  author       = {{Google DeepMind}},
  year         = {2025},
  howpublished = {\url{https://deepmind.google/models/gemini/flash/}}
}

@misc{GROBID,
    title = {GROBID},
    howpublished = {\url{https://github.com/kermitt2/grobid}},
    publisher = {GitHub},
    year = {2008--2026}
}

@misc{neo4j_platform,
  title        = {Neo4j Graph Database \& Analytics Platform},
  author       = {{Neo4j, Inc.}},
  year         = {2026},
  howpublished = {\url{https://neo4j.com/}}
}

@misc{chen2025m3embeddingmultilingualitymultifunctionalitymultigranularity,
      title={M3-Embedding: Multi-Linguality, Multi-Functionality, Multi-Granularity Text Embeddings Through Self-Knowledge Distillation}, 
      author={Jianlv Chen and Shitao Xiao and Peitian Zhang and Kun Luo and Defu Lian and Zheng Liu},
      year={2025},
      eprint={2402.03216},
      archivePrefix={arXiv},
      primaryClass={cs.CL},
      url={https://arxiv.org/abs/2402.03216}, 
}

@misc{mcp_anthropic,
  title        = {Model Context Protocol (MCP)},
  author       = {{Anthropic}},
  year         = {2024},
  howpublished = {\url{https://modelcontextprotocol.io/docs/getting-started/intro}}
}

@misc{a2a_protocol,
  title        = {Agent-to-Agent (A2A) Protocol},
  author       = {{Google}},
  year         = {2024},
  howpublished = {\url{https://github.com/google/A2A}},
}

@misc{falcon_outage,
author = {},
year={2022},
title = {{Falcon Cable Fault Believed To Be From Air Strike}},
howpublished = {\url{https://subtelforum.com/falcon-cable-fault-believed-to-be-from-air-strike/}},
}

@misc{gatechIODA,
	author = {},
	title = {{I}{O}{D}{A} --- ioda.inetintel.cc.gatech.edu},
	howpublished = {\url{https://ioda.inetintel.cc.gatech.edu}},
	year = {2025},
}

@misc{cloudflare,
	author = {},
	title = {{W}orldwide {O}verview | {C}loudflare {R}adar --- radar.cloudflare.com},
	howpublished = {\url{https://radar.cloudflare.com}},
	year = {2025},
}

@misc{netblocks,
	author = {},
	title = {{H}ome - {N}et{B}locks --- netblocks.org},
	howpublished = {\url{https://netblocks.org}},
	year = {2025},
}

@misc{routeviews,
	author = {},
	title = {{R}oute{V}iews; {U}niversity of {O}regon {R}oute{V}iews {P}roject --- routeviews.org},
	howpublished = {\url{https://www.routeviews.org/routeviews/}},
	year = {2025},
}

@misc{riperis,
	author = {},
	title = {{R}outing {I}nformation {S}ervice ({R}{I}{S}) --- ripe.net},
	howpublished = {\url{https://www.ripe.net/analyse/internet-measurements/routing-information-service-ris/}},
	year = {2025},
}

@misc{bgptools,
	author = {},
	title = {{B}{G}{P}.{T}ools --- bgp.tools},
	howpublished = {\url{https://bgp.tools}},
	year = {2025},
}

@inproceedings{bgpstream,
author = {Orsini, Chiara and King, Alistair and Giordano, Danilo and Giotsas, Vasileios and Dainotti, Alberto},
title = {BGPStream: A Software Framework for Live and Historical BGP Data Analysis},
year = {2016},
isbn = {9781450345262},
publisher = {Association for Computing Machinery},
address = {New York, NY, USA},
url = {https://doi.org/10.1145/2987443.2987482},
doi = {10.1145/2987443.2987482},
booktitle = {Proceedings of the 2016 Internet Measurement Conference},
pages = {429–444},
numpages = {16},
location = {Santa Monica, California, USA},
series = {IMC '16}
}

@inproceedings{bgpdetecting,
  title={Detecting BGP configuration faults with static analysis},
  author={Feamster, Nick and Balakrishnan, Hari},
  booktitle={Proceedings of the 2nd conference on Symposium on Networked Systems Design \& Implementation-Volume 2},
  pages={43--56},
  year={2005}
}

@article{bgpstate,
  title={The state of the art in bgp visualization tools: A mapping of visualization techniques to cyberattack types},
  author={Raynor, Justin and Crnovrsanin, Tarik and Di Bartolomeo, Sara and South, Laura and Saffo, David and Dunne, Cody},
  journal={IEEE Transactions on Visualization and Computer Graphics},
  volume={29},
  number={1},
  pages={1059--1069},
  year={2022},
  publisher={IEEE}
}

@inproceedings{bgpnext,
  title={The Next Generation of BGP Data Collection Platforms},
  author={Alfroy, Thomas and Holterbach, Thomas and Krenc, Thomas and Claffy, KC and Pelsser, Cristel},
  booktitle={Proceedings of the ACM SIGCOMM 2024 Conference},
  pages={794--812},
  year={2024}
}

@article{albgp,
  title={BGP Anomaly Detection Techniques: A Survey},
  author={Al-Musawi, Bahaa and Branch, Philip and Armitage, Grenville},
  journal={IEEE Communications Surveys \& Tutorials},
  volume={19},
  number={1},
  pages={377--396},
  year={2016},
  publisher={IEEE}
}

@inproceedings{tracerouteinternet,
  title={Internet scale reverse traceroute},
  author={Vermeulen, Kevin and Gurmericliler, Ege and Cunha, Italo and Choffnes, David and Katz-Bassett, Ethan},
  booktitle={Proceedings of the 22nd ACM Internet Measurement Conference},
  pages={694--715},
  year={2022}
}

@inproceedings{traceroutemao,
  title={Towards an accurate AS-level traceroute tool},
  author={Mao, Zhuoqing Morley and Rexford, Jennifer and Wang, Jia and Katz, Randy H},
  booktitle={Proceedings of the 2003 conference on Applications, technologies, architectures, and protocols for computer communications},
  pages={365--378},
  year={2003}
}

@inproceedings{paristraceroute,
  title={Avoiding traceroute anomalies with Paris traceroute},
  author={Augustin, Brice and Cuvellier, Xavier and Orgogozo, Benjamin and Viger, Fabien and Friedman, Timur and Latapy, Matthieu and Magnien, Cl{\'e}mence and Teixeira, Renata},
  booktitle={Proceedings of the 6th ACM SIGCOMM conference on Internet measurement},
  pages={153--158},
  year={2006}
}

@inproceedings{traceroutemultilevel,
  title={Multilevel MDA-lite Paris traceroute},
  author={Vermeulen, Kevin and Strowes, Stephen D and Fourmaux, Olivier and Friedman, Timur},
  booktitle={Proceedings of the Internet Measurement Conference 2018},
  pages={29--42},
  year={2018}
}

@article{topologyzoo,
  title={The internet topology zoo},
  author={Knight, Simon and Nguyen, Hung X and Falkner, Nickolas and Bowden, Rhys and Roughan, Matthew},
  journal={IEEE Journal on Selected Areas in Communications},
  volume={29},
  number={9},
  pages={1765--1775},
  year={2011},
  publisher={IEEE}
}

@article{topologycollecting,
  title={Collecting the Internet AS-level topology},
  author={Zhang, Beichuan and Liu, Raymond and Massey, Daniel and Zhang, Lixia},
  journal={ACM SIGCOMM Computer Communication Review},
  volume={35},
  number={1},
  pages={53--61},
  year={2005},
  publisher={ACM New York, NY, USA}
}

@article{perfsurvey,
  title={A survey on internet performance measurement platforms and related standardization efforts},
  author={Bajpai, Vaibhav and Sch{\"o}nw{\"a}lder, J{\"u}rgen},
  journal={IEEE Communications Surveys \& Tutorials},
  volume={17},
  number={3},
  pages={1313--1341},
  year={2015},
  publisher={IEEE}
}

@article{latlong,
  title={LatLong: Diagnosing wide-area latency changes for CDNs},
  author={Zhu, Yaping and Helsley, Benjamin and Rexford, Jennifer and Siganporia, Aspi and Srinivasan, Sridhar},
  journal={IEEE Transactions on Network and Service Management},
  volume={9},
  number={3},
  pages={333--345},
  year={2012},
  publisher={IEEE}
}

@inproceedings{cdnmoving,
  title={Moving beyond end-to-end path information to optimize CDN performance},
  author={Krishnan, Rupa and Madhyastha, Harsha V and Srinivasan, Sridhar and Jain, Sushant and Krishnamurthy, Arvind and Anderson, Thomas and Gao, Jie},
  booktitle={Proceedings of the 9th ACM SIGCOMM conference on Internet measurement},
  pages={190--201},
  year={2009}
}

@inproceedings{cdnserver,
  title={A server-to-server view of the Internet},
  author={Chandrasekaran, Balakrishnan and Smaragdakis, Georgios and Berger, Arthur and Luckie, Matthew and Ng, Keung-Chi},
  booktitle={Proceedings of the 11th ACM Conference on Emerging Networking Experiments and Technologies},
  pages={1--13},
  year={2015}
}

@misc{2024codellamaopenfoundation,
      title={Code Llama: Open Foundation Models for Code}, 
      author={Baptiste Rozière and Jonas Gehring and Fabian Gloeckle and Sten Sootla and Itai Gat and Xiaoqing Ellen Tan and Yossi Adi and Jingyu Liu and Romain Sauvestre and Tal Remez and Jérémy Rapin and Artyom Kozhevnikov and Ivan Evtimov and Joanna Bitton and Manish Bhatt and Cristian Canton Ferrer and Aaron Grattafiori and Wenhan Xiong and Alexandre Défossez and Jade Copet and Faisal Azhar and Hugo Touvron and Louis Martin and Nicolas Usunier and Thomas Scialom and Gabriel Synnaeve},
      year={2024},
      eprint={2308.12950},
      archivePrefix={arXiv},
      primaryClass={cs.CL},
      url={https://arxiv.org/abs/2308.12950}, 
}

@misc{wei2023chainofthoughtpromptingelicitsreasoning,
      title={Chain-of-Thought Prompting Elicits Reasoning in Large Language Models}, 
      author={Jason Wei and Xuezhi Wang and Dale Schuurmans and Maarten Bosma and Brian Ichter and Fei Xia and Ed Chi and Quoc Le and Denny Zhou},
      year={2023},
      eprint={2201.11903},
      archivePrefix={arXiv},
      primaryClass={cs.CL},
      url={https://arxiv.org/abs/2201.11903}, 
}

@article{Wang_2024,
   title={A survey on large language model based autonomous agents},
   volume={18},
   ISSN={2095-2236},
   url={http://dx.doi.org/10.1007/s11704-024-40231-1},
   DOI={10.1007/s11704-024-40231-1},
   number={6},
   journal={Frontiers of Computer Science},
   publisher={Springer Science and Business Media LLC},
   author={Wang, Lei and Ma, Chen and Feng, Xueyang and Zhang, Zeyu and Yang, Hao and Zhang, Jingsen and Chen, Zhiyuan and Tang, Jiakai and Chen, Xu and Lin, Yankai and Zhao, Wayne Xin and Wei, Zhewei and Wen, Jirong},
   year={2024},
   month=mar }

@misc{adrs,
      title={Barbarians at the Gate: How AI is Upending Systems Research}, 
      author={Audrey Cheng and Shu Liu and Melissa Pan and Zhifei Li and Bowen Wang and Alex Krentsel and Tian Xia and Mert Cemri and Jongseok Park and Shuo Yang and Jeff Chen and Lakshya Agrawal and Aditya Desai and Jiarong Xing and Koushik Sen and Matei Zaharia and Ion Stoica},
      year={2025},
      eprint={2510.06189},
      archivePrefix={arXiv},
      primaryClass={cs.AI},
      url={https://arxiv.org/abs/2510.06189}, 
}

@misc{glia,
      title={Glia: A Human-Inspired AI for Automated Systems Design and Optimization}, 
      author={Pouya Hamadanian and Pantea Karimi and Arash Nasr-Esfahany and Kimia Noorbakhsh and Joseph Chandler and Ali ParandehGheibi and Mohammad Alizadeh and Hari Balakrishnan},
      year={2025},
      eprint={2510.27176},
      archivePrefix={arXiv},
      primaryClass={cs.AI},
      url={https://arxiv.org/abs/2510.27176}, 
}

@inproceedings{popsicle,
author = {Durairajan, Ramakrishnan and Sommers, Joel and Barford, Paul},
title = {Layer 1-informed Internet Topology Measurement},
year = {2014},
isbn = {9781450332132},
publisher = {Association for Computing Machinery},
address = {New York, NY, USA},
url = {https://doi.org/10.1145/2663716.2663737},
doi = {10.1145/2663716.2663737},
pages = {381–394},
numpages = {14},
location = {Vancouver, BC, Canada},
series = {IMC '14}
}

@inproceedings{syslogdigest
,
author = {Qiu, Tongqing and Ge, Zihui and Pei, Dan and Wang, Jia and Xu, Jun},
title = {What happened in my network: mining network events from router syslogs},
year = {2010},
isbn = {9781450304832},
publisher = {Association for Computing Machinery},
address = {New York, NY, USA},
url = {https://doi.org/10.1145/1879141.1879202},
doi = {10.1145/1879141.1879202},
pages = {472–484},
numpages = {13},
location = {Melbourne, Australia},
series = {IMC '10}
}

@article{tessellation,
author = {Xia, Ning and Song, Han Hee and Liao, Yong and Iliofotou, Marios and Nucci, Antonio and Zhang, Zhi-Li and Kuzmanovic, Aleksandar},
title = {Mosaic: quantifying privacy leakage in mobile networks},
year = {2013},
issue_date = {October 2013},
publisher = {Association for Computing Machinery},
address = {New York, NY, USA},
volume = {43},
number = {4},
issn = {0146-4833},
url = {https://doi.org/10.1145/2534169.2486008},
doi = {10.1145/2534169.2486008},
journal = {SIGCOMM Comput. Commun. Rev.},
month = aug,
pages = {279–290},
numpages = {12}
}

@inproceedings{prefix2org,
author = {Gouda, Deepak and Dainotti, Alberto and Testart, Cecilia},
title = {Prefix2Org: Mapping BGP Prefixes to Organizations},
year = {2025},
isbn = {9798400718601},
publisher = {Association for Computing Machinery},
address = {New York, NY, USA},
url = {https://doi.org/10.1145/3730567.3764485},
doi = {10.1145/3730567.3764485},
pages = {397–414},
numpages = {18},
location = {USA},
series = {IMC '25}
}

@misc{submarine_map,
author = {},
  title = {Submarine Cable Map},
  url = "https://www.submarinecablemap.com/",
month = {},
year = {2025}
}

@inproceedings{arachnet,
author = {Ramanathan, Alagappan and Kang, Eunju and Han, Dongsu and Abdu Jyothi, Sangeetha},
title = {Towards an Agentic Workflow for Internet Measurement Research},
year = {2025},
isbn = {9798400722806},
publisher = {Association for Computing Machinery},
address = {New York, NY, USA},
url = {https://doi.org/10.1145/3772356.3772409},
doi = {10.1145/3772356.3772409},
pages = {61–68},
numpages = {8},
location = {UMD Campus, College Park, MD, USA},
series = {HotNets '25}
}

\clearpage
\appendix
\appendix

\label{sec:appendix:cost}
\begin{table*}[h]
\centering
\begin{tabular}{@{}p{4cm}cp{6cm}@{}}
\toprule
\textbf{Component} & \textbf{Cost per Run} & \textbf{Notes} \\
\midrule
Agent 1 (QueryMind) & \$0.06 - \$0.15 & Query decomposition \\
Agent 2 (WorkflowScout) & \$0.25 - \$0.60 & Workflow specification \\
Agent 3 (SolutionWeaver) & \$0.35 - \$0.60 & Code generation \\
Agent 4 (RegistryCurator) & \$0.20 - \$0.40 & Quality assessment \\[0.15em]
\textbf{Standard Workflow} & \textbf{\$0.80 - \$1.70} & \textbf{Agents 1-3 only} \\[0.15em]
Verification Synthesizer & \$0.50 - \$0.70 & Standard mode \\
Verification Synthesizer & \$1.00 - \$1.50 & With comparison \\
12-Variant Generation & \$2.50 - \$3.00 & Model comparison \\[0.15em]
Validation Strategizer & \$0.16 - \$0.30 & Strategy generation \\
Validation CodeGenerator & \$0.40 - \$0.60 & Validation code \\[0.15em]
\textbf{Full Evaluation} & \textbf{\$4.50 - \$7.00} & \textbf{12 variants + verification + validation} \\
\bottomrule
\end{tabular}
\caption{Cost Analysis Summary}
\label{tab:cost_analysis}
\end{table*}

\section{Cost Analysis}
\label{sec:cost-analysis}

All agents in Airavat's experiments use Claude models (Sonnet 4.5 and Opus 4.5) with the following costs estimated per execution in Table~\ref{tab:cost_analysis}. The Multi-Agent Workflow Generation pipeline comprises four agents with varying computational requirements. Agent 1 (QueryMind) costs \$0.06-\$0.15 per run for problem characterization and knowledge graph query formulation. Agent 2 (WorkflowScout) costs \$0.25-\$0.60 per run for workflow design generation, reflecting the more complex reasoning required to synthesize measurement literature into concrete specifications. Agent 3 (SolutionWeaver) costs \$0.35-\$0.60 per run for code generation from workflow specifications. Agent 4 (RegistryCurator) costs \$0.20-\$0.40 per run when evaluating generated workflows against literature-derived quality criteria.

The Verification Engine adds substantial computational overhead for quality assurance. The Synthesizer agent costs \$0.50-\$0.70 per run in standard mode, or \$1.00-\$1.50 when performing best-approach comparison with the synthesized workflow. Generating 12 workflow variants (4 models × 3 temperatures) for comprehensive model comparison costs approximately \$2.50-\$3.00 per run. The Validation Engine comprises two agents: the Strategizer costs \$0.16-\$0.30 per run for generating validation strategies from measurement literature, while the CodeGenerator costs \$0.40-\$0.60 per run for translating strategies into executable validation code.

Estimating total costs for all case studies presented in the paper yields a modest overall expenditure. Case Studies 1-3 each required one standard workflow generation run, totaling approximately \$3.75 (3 × \$1.25 midpoint). Case Study 4 employed comprehensive verification with 3 independent runs of 12 workflow variants each, costing approximately \$17.25 (3 × \$5.75 midpoint). The Validation Engine demonstration required strategy generation and code generation, adding approximately \$0.45. The Knowledge Graph extraction evaluation used only local LLMs (LLama-3.1-8B and LLama3.3-70B), incurring negligible cloud API costs. Total estimated expenditure for all experiments presented is approximately \$21.50, demonstrating the cost-effectiveness of the approach for research-scale evaluation.

Cost optimization represents a significant opportunity for future work. Very little optimization was performed in the current implementation—intermediate outputs from earlier pipeline stages were often passed in entirety to subsequent agents rather than extracting only relevant content. For example, Agent 2 (WorkflowScout) receives complete Agent 1 (QueryMind) results rather than filtered excerpts. Selective content extraction could reduce token consumption by an estimated 30-40\%, lowering per-run costs to \$0.50-\$1.00 for standard workflows and \$3.00-\$4.50 for full evaluation runs. Additional optimizations include caching repeated knowledge graph queries, compressing intermediate representations, and employing smaller models for routine validation checks while reserving powerful models for complex reasoning tasks. These improvements would enhance system affordability while maintaining generation quality, making Airavat more accessible for broader research community adoption.

\section{Knowledge Graph Extraction Details and Quality} 
\label{sec:kg-details}

Airavat constructs a Neo4j knowledge graph encoding measurement domain knowledge through semantic embeddings and typed relationships. The entity model includes ten entity types: Papers, Problems, ResearchGaps, Approaches, PipelineSteps, Algorithms, Metrics, Parameters, Datasets, and Validations. Entities connect through typed relationships. Papers PROPOSE Approaches that SOLVE Problems; Approaches USE\_DATASET and are VALIDATED\_BY validation methodologies; Approaches contain ordered PIPELINE\_STEP sequences that USE\_ALGORITHM; IMPROVES\_UPON relationships capture methodological evolution. This schema enables similarity-based retrieval, relationship traversal for methodology evolution, and constraint-based filtering. Airavat uses MD5 hashing for deduplication and BGE-M3 embeddings for semantic search.

To evaluate the extraction quality of our local LLM approach (LLama-3.1-8B and LLama3.3-70B), we conducted a validation study using Claude Sonnet 4.5 as an evaluator. We selected seven representative papers covering various Internet measurement topics (submarine cable mapping, network resilience analysis, routing communities, broadband availability, third-party dependencies, and IPv6 allocation) and compared the extraction outputs from our local LLM pipeline against what Sonnet itself would have generated. We evaluated overlap across the same five key extraction categories: problem statement, methodology, datasets, baselines, and validations. Across the seven papers, the local LLM extraction achieved an average aggregate overlap of 87.3\% with Sonnet's extraction, with individual paper scores ranging from 72\% to 94\%. This demonstrates that cost-effective local LLM extraction achieves strong performance comparable to expensive cloud-based models for populating the knowledge graph, validating our design choice to use local models for large-scale paper processing while reserving cloud LLMs for the agent pipeline's reasoning tasks.

\section{Knowledge Graph Extraction Example}

We demonstrate the extraction pipeline's output using the Nautilus paper~\cite{nautilus} as a representative example (Figure~\ref{fig:kg_output},~\ref{fig:kg_output_2}). The extraction captures five key categories from measurement papers. Note that the following is a condensed version for illustration purposes—the actual extraction contains more comprehensive details for each section.

\parab{Extraction Structure} The knowledge graph extraction system processes papers to extract structured information across five primary categories: Problem Statement, Methodology, Datasets, Baseline Comparisons, and Validations.

\parab{Extraction Characteristics} This condensed example in Figure~\ref{fig:kg_output},~\ref{fig:kg_output_2} illustrates the structured extraction approach applied across all papers in the knowledge graph. The full extraction contains comprehensive details, including all 10 methodology steps, 19 data sources with detailed characteristics, complete baseline comparisons, and extensive validation experiments. The extraction enables the knowledge graph to answer queries like "What clustering algorithms are used for geolocation?" (DBSCAN with $\epsilon$=20km), "What validation strategies were employed?" (cable failures, targeted measurements, operator maps), and "What were the key parameters?" (SoL threshold=0.05, radius=500km, weights 5:4:1).

\section{Knowledge Graph Characteristics} ~\label{kg_characteristic}

The knowledge graph constructed for Airavat aggregates structured information from Internet measurement research papers. The graph represents a comprehensive corpus of measurement methodologies, techniques, and validation approaches extracted from published literature. Below, we provide the scale and composition of the knowledge graph to demonstrate its breadth and depth for supporting workflow generation.

The knowledge graph contains 2,021 research papers processed through the extraction pipeline. From these papers, the system extracted 1,944 distinct problem statements characterizing measurement challenges, and 3,767 unique approaches describing solution methodologies. The methodology extraction identified 13,813 pipeline steps representing the granular procedural decomposition of measurement workflows. Technical components include 2,956 algorithms (clustering methods, graph traversal techniques, optimization procedures), 7,018 datasets and data sources (measurement platforms, geolocation services, routing databases), and 4,483 parameters (thresholds, weights, configuration values). Validation information comprises 722 validation strategies and 1,016 evaluation metrics extracted from experimental sections.

The knowledge graph maintains 65,250 relationships connecting these entities, enabling traversal between related concepts. These relationships link problems to applicable approaches, approaches to constituent pipeline steps, steps to required algorithms and datasets, and methodologies to validation strategies. The relationship structure enables queries like "What validation strategies were used for geolocation-based approaches?" or "What datasets are required for cross-layer mapping problems?" to return contextually relevant results grounded in measurement literature. This scale demonstrates that the knowledge graph provides substantial coverage of Internet measurement research, capturing diverse problem domains (infrastructure resilience, prefix mapping, network topology inference), methodological techniques (geolocation clustering, BGP analysis, traceroute processing), and validation practices (cable failure analysis, targeted measurements, ground truth comparison). The comprehensive representation enables Airavat agents to reason about measurement problems through literature grounded context rather than relying solely on LLM parametric knowledge.

\begin{figure*}
\begin{lstlisting}[style=jsonstyle]

  (*@\textbf{\textcolor{blue}{"paper\_title"}}@*): "Nautilus: Framework for Cross-Layer Cartography of Submarine Cables and IP Links",
  
  (*@\textbf{\textcolor{blue}{"extractions"}}@*): {
  
    (*@\textbf{"Problem Statement"}@*):
        {
          "problem_statement": "Mapping IP links to submarine cables accurately",
          
          "problem_details": {
                "what_is_lacking":"Existing approaches lack accuracy and are coarse-grained",
                "why_its_challenging":"Geolocation inaccuracies, incomplete ownership information, complex topology",
                "scope":"At Internet scale, for critical infrastructure"},
          
          "research_gaps": [
                "Accurate cross-layer mapping",
                "Handling geolocation uncertainties and incomplete data" ] 
        },
        
    (*@\textbf{"Methodology"}@*): 
        {
          "approach_overview": "Nautilus uses publicly available datasets and techniques to generate IP link to submarine cable mapping with confidence scores.",
          "pipeline": [
                ...
                {
                  "step_number": 2,
                  "step_name": "Geolocation Module",
                  "description": "Collect and aggregate geolocation from eleven services, classify IP links",
                  "algorithms_used": ["DBSCAN"],
                  "parameters": {
                        "minPoints": "1",
                        "epsilon": "20 km" } }, 
                ...
                {
                  "step_number": 10,
                  "step_name": "Aggregation & Final Mapping",
                  "description": "Combine geolocation and cable owner outputs",
                  "parameters": {
                        "weightage": "0.5 geolocation, 0.4 distance, 0.1 ownership" }}
            ],
          "algorithms": [ {
                  "name": "DBSCAN",
                  "purpose": "Clustering geolocations based on density",
                  "parameters": {
                        "minPoints": "1",
                        "epsilon": "20 km" }} ] 
        },
\end{lstlisting}
\caption{A representative subset of the extraction for the Nautilus paper on submarine cable mapping. Continued in Figure~\ref{fig:kg_output_2}}
\label{fig:kg_output}
\end{figure*}

\begin{figure*}
\begin{lstlisting}[style=jsonstyle]

    (*@\textbf{"Datasets or Data Sources"}@*):
        {
          "data_sources": [
                {
                  "name": "RIPE Atlas",
                  "what_it_contains": "Traceroute data",
                  "how_used": "Collecting traceroutes for IP link extraction",
                  "characteristics": {
                        "size": "~120M traceroutes" } },
                {
                  "name": "Telegeography map",
                  "what_it_contains": "Submarine cable information",
                  "how_used": "Mapping IP links to submarine cables",
                  "characteristics": {
                        "size": "~480 submarine cables" } } ],
          ...
        },
        
    (*@\textbf{"Baseline Comparisons"}@*):
        {
          "baselines": [
                {
                  "baseline_name": "SCN-Crit",
                  "baseline_approach": "Uses drivability metric, maps cables at country level",
                  "comparison_metrics": [ {
                          "metric": "Number of cables predicted per link",
                          "improvement": "Nautilus predicts 35% fewer cables per link" } ],
                  "baseline_limitations": [
                        "Maps cables at country level",
                        "Conservative approach misses potential submarine paths" ] } ]
        },
        
    (*@\textbf{"Validations"}@*):
        {
          "validations": [
                ...
                {
                  "validation_name": "Submarine Cable Failures",
                  "methodology": "Analyze IP link disappearance during documented cable failures",
                  "specific_examples": [ {
                      "example_name": "Yemen Outage Analysis",
                      "results": { "disappearance_of_links": "106 links disappeared" } } ] },
                {
                  "validation_name": "Targeted Traceroutes",
                  "methodology": "Targeted measurements between RIPE probes near cable landing points",
                  "results": { "match_rate": "77% top prediction match" } },
                ... 
                ]
        } 
\end{lstlisting}
\caption{Continued from Figure~\ref{fig:kg_output} -- The representative subset of the extraction for the Nautilus paper on submarine cable mapping.}
\label{fig:kg_output_2}
\end{figure*}

\begin{figure*}
\begin{lstlisting}[style=jsonstyle]
  (*@\textbf{"query\_summary"}@*): "Concise 1-2 sentence summary of what user is asking",
  (*@\textbf{"complexity\_assessment"}@*): {{
        "temporal":     {{  "answer": "yes/no",
                            "reasoning": "Explanation" }},
        "spatial":      {{  "answer": "yes/no",
                            "reasoning": "Explanation" }},
        "causal":       {{  "answer": "yes/no",
                            "reasoning": "Explanation" }},
        "stakeholder":  {{  "answer": "yes/no",
                            "reasoning": "Explanation" }},
        "data":         {{  "answer": "yes/no",
                            "reasoning": "Explanation" }},
        "score": 0,
        "tier": "simple/moderate/complex",
        "rationale": "Overall complexity justification",
        "implications_for_agent2": "What this complexity means for solution design."}},
  
  (*@\textbf{"sub\_problems"}@*): [ {{    
        "id": "SP1",
        "description": "Detailed sub-problem description",
        "dependencies": ["SP2", "SP3"],
        "priority": "high/medium/low",
        "estimated_difficulty": "Description of difficulty" }} ], 
        
  (*@\textbf{"constraints"}@*): {{
        "technical": ["constraint 1", "constraint 2"],
        "data": ["constraint 1", "constraint 2"],
        "methodological": ["constraint 1", "constraint 2"],
        "temporal": ["constraint 1", "constraint 2"] }},
        
  (*@\textbf{"success\_criteria"}@*): {{
        "primary": "Main success criterion",
        "secondary": ["criterion 1", "criterion 2"],
        "validation_approach": "How to validate success" }},
        
  (*@\textbf{"risks"}@*): [ {{
        "risk": "Description of risk",
        "likelihood": "high/medium/low",
        "severity": "high/medium/low",
        "mitigation": "How to mitigate" }}],
        
  (*@\textbf{"registry\_mapping"}@*): {{
        "relevant_functions": [ {{
                "function_name": "name_from_registry",
                "purpose": "What it provides for this query",
                "sub_problems_addressed": ["SP1", "SP2"] }}],
        "integration_points": ["How functions connect"],
        "gaps": ["What registry doesn't provide"] }},
        
  (*@\textbf{"recommendations\_for\_designer"}@*): [
        "Recommendation 1 for Agent 2",
        "Recommendation 2 for Agent 2" ]
\end{lstlisting}
\caption{\small Output of \textit{QueryMind} follows this schema to examine complexities of subproblems, evaluate constraints, define success criteria, and map to the relevant registry to provide guidance to \textit{WorkflowScout}.}
\label{fig:querymind_schema}
\end{figure*}

\begin{table*}[t]
\centering
\begin{tabular}{@{}lp{8cm}@{}}
\toprule
\textbf{Case Study} & \textbf{Registry Functions by System} \\
\midrule
\textbf{Cable Impact} & 
\textbf{nautilus\_system:} 
\begin{enumerate}[leftmargin=*,nosep]
  \item get\_nautilus\_link\_to\_cable\_mapping
  \item get\_lp\_id\_to\_country\_dict
  \item get\_ip\_to\_geolocation\_mappings
  \item get\_ip\_to\_asn\_mappings
\end{enumerate} \\[1.5em]

\textbf{Disaster Impact} & 
\textbf{nautilus\_system:} 
\begin{enumerate}[leftmargin=*,nosep]
  \item get\_nautilus\_link\_to\_cable\_mapping
  \item get\_lp\_id\_to\_country\_dict
  \item get\_ip\_to\_geolocation\_mappings
  \item get\_ip\_to\_asn\_mappings
\end{enumerate}
\textbf{xaminer\_system:}
\begin{enumerate}[leftmargin=*,nosep,start=5]
  \item generate\_cable\_segments\_to\_all\_info\_map
  \item generate\_cable\_segment\_to\_country\_as\_maps
  \item process\_single\_event
\end{enumerate} \\[1.5em]

\textbf{Cascading Effects} & 
\textbf{nautilus\_system:} 
\begin{enumerate}[leftmargin=*,nosep]
  \item get\_nautilus\_link\_to\_cable\_mapping
  \item get\_lp\_id\_to\_country\_dict
  \item get\_ip\_to\_geolocation\_mappings
  \item get\_ip\_to\_asn\_mappings
\end{enumerate}
\textbf{xaminer\_system:}
\begin{enumerate}[leftmargin=*,nosep,start=5]
  \item generate\_cable\_segments\_to\_all\_info\_map
  \item generate\_cable\_segment\_to\_country\_as\_maps
  \item process\_single\_event
\end{enumerate}
\textbf{submarine\_system:}
\begin{enumerate}[leftmargin=*,nosep,start=8]
  \item get\_country\_to\_cable\_graph
\end{enumerate}
\textbf{as\_dependency\_system:}
\begin{enumerate}[leftmargin=*,nosep,start=9]
  \item get\_as\_dependency\_graph
\end{enumerate} \\[1.5em]

\textbf{Prefix2Org} & 
\textbf{bgp\_system:} 
\begin{enumerate}[leftmargin=*,nosep]
  \item download\_bgp\_dumps
\end{enumerate}
\textbf{whois\_system:}
\begin{enumerate}[leftmargin=*,nosep,start=2]
  \item parse\_whois\_dump
\end{enumerate}
\textbf{rpki\_system:}
\begin{enumerate}[leftmargin=*,nosep,start=3]
  \item get\_rpki\_snapshot
\end{enumerate}
\textbf{as2org\_system:}
\begin{enumerate}[leftmargin=*,nosep,start=4]
  \item get\_as2org\_mappings
\end{enumerate} \\
\bottomrule
\end{tabular}
\caption{Registry functions utilized across case studies.}
\label{tab:registry_functions}
\end{table*}

\begin{table*}[t]
\centering
\begin{tabular}{@{}p{4.5cm}cp{3.5cm}p{6.5cm}@{}}
\toprule
\textbf{Parameter} & \textbf{Value} & \textbf{Component} & \textbf{Purpose} \\
\midrule
Novelty threshold (high) & 0.7 & Evaluator - Stage 2 & Problems above classified as novel \\[0.4em]
Novelty threshold (low) & 0.4 & Evaluator - Stage 2 & Problems below classified as well-studied \\[0.4em]
Infeasibility threshold & 60 & Evaluator - Stage 2 & Sub-dimension scores below this trigger rejection \\[0.4em]
Excellence threshold & 85 & Selector & Workflows above approved directly \\[0.4em]
Good range lower bound & 80 & Selector & Workflows below always require synthesis \\[0.4em]
Structural diversity threshold & 0.4 & Selector & Triggers complementarity analysis \\[0.4em]
Weight adjustment bound & $\pm$0.2 & Evaluator - Stage 3 & Maximum adjustment per dimension \\
\bottomrule
\end{tabular}
\small
\caption{Verification Engine parameters.}
\label{tab:parameters}
\end{table*}

\begin{figure*}[ht]
\centering
\includegraphics[width=0.9\textwidth]{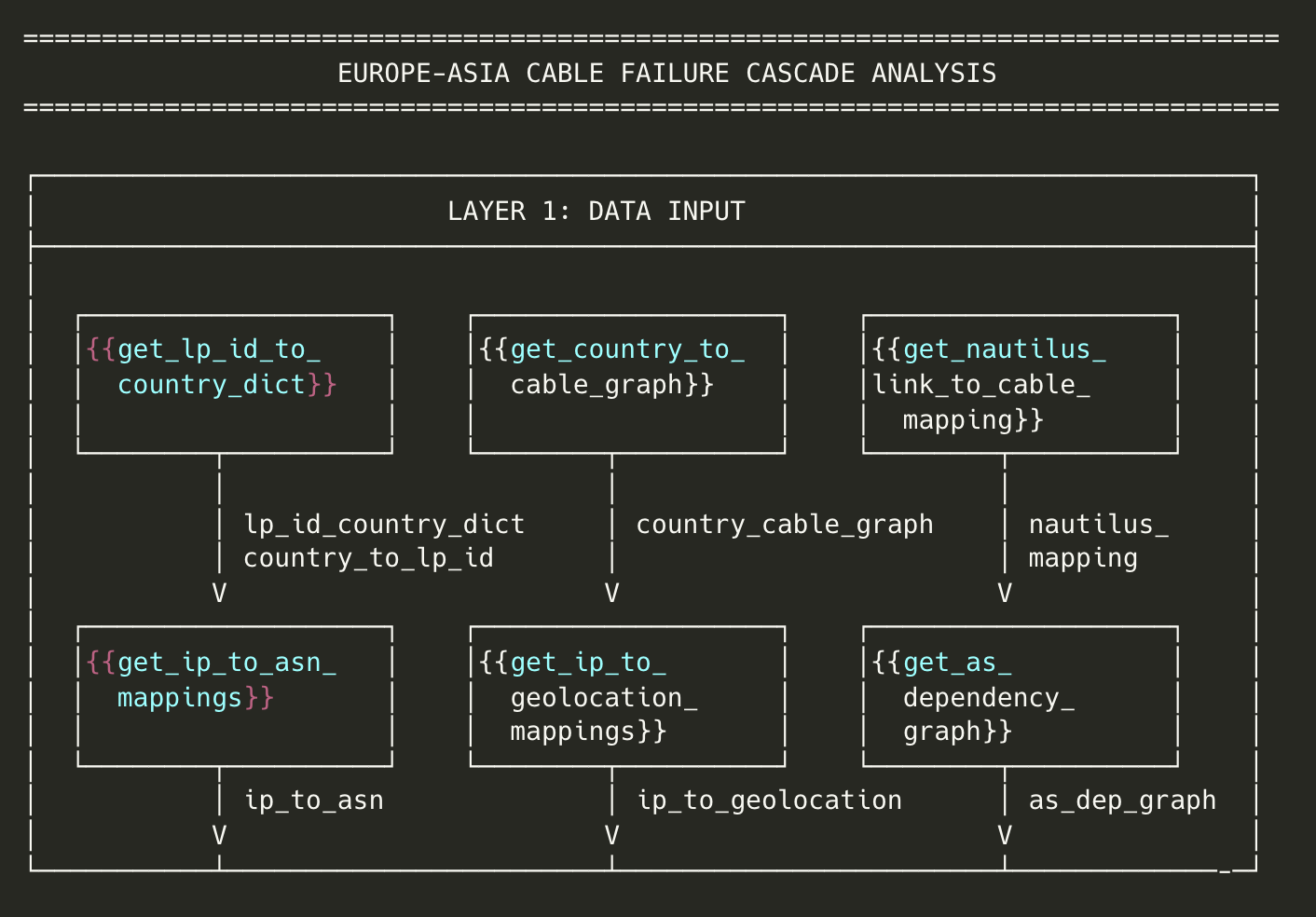}
\caption{Europe-Asia Cable Failure Cascade Analysis Workflow (Data Input Layer)}
\label{fig:cascade_workflow1}
\end{figure*}

\begin{figure*}[p]
\centering
\includegraphics[width=\textwidth,height=0.9\textheight,keepaspectratio]{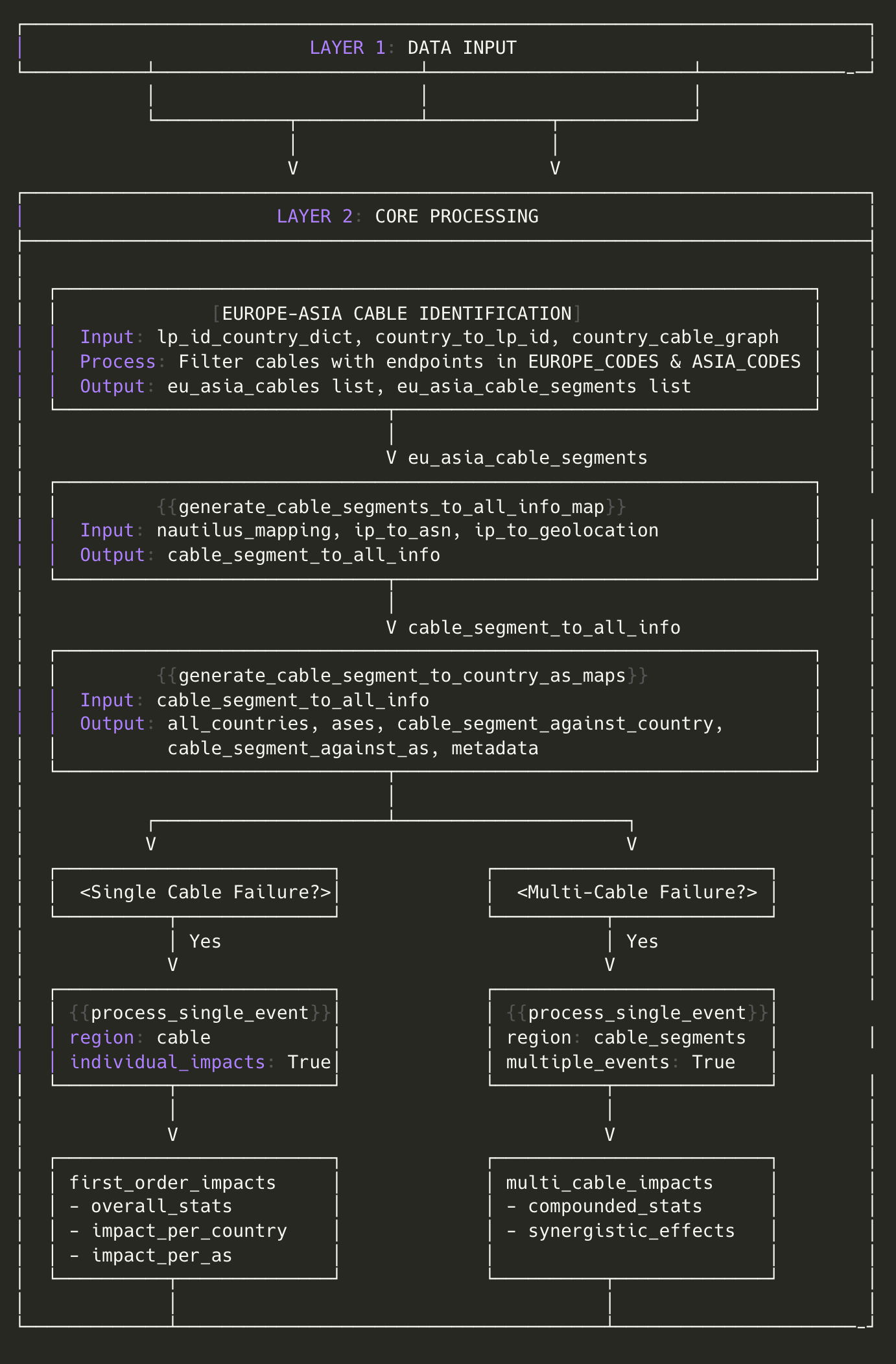}
\caption{Europe-Asia Cable Failure Cascade Analysis Workflow (Core Processing Layer)}
\label{fig:cascade_workflow2}
\end{figure*}

\begin{figure*}[p]
\centering
\includegraphics[width=0.8\textwidth]{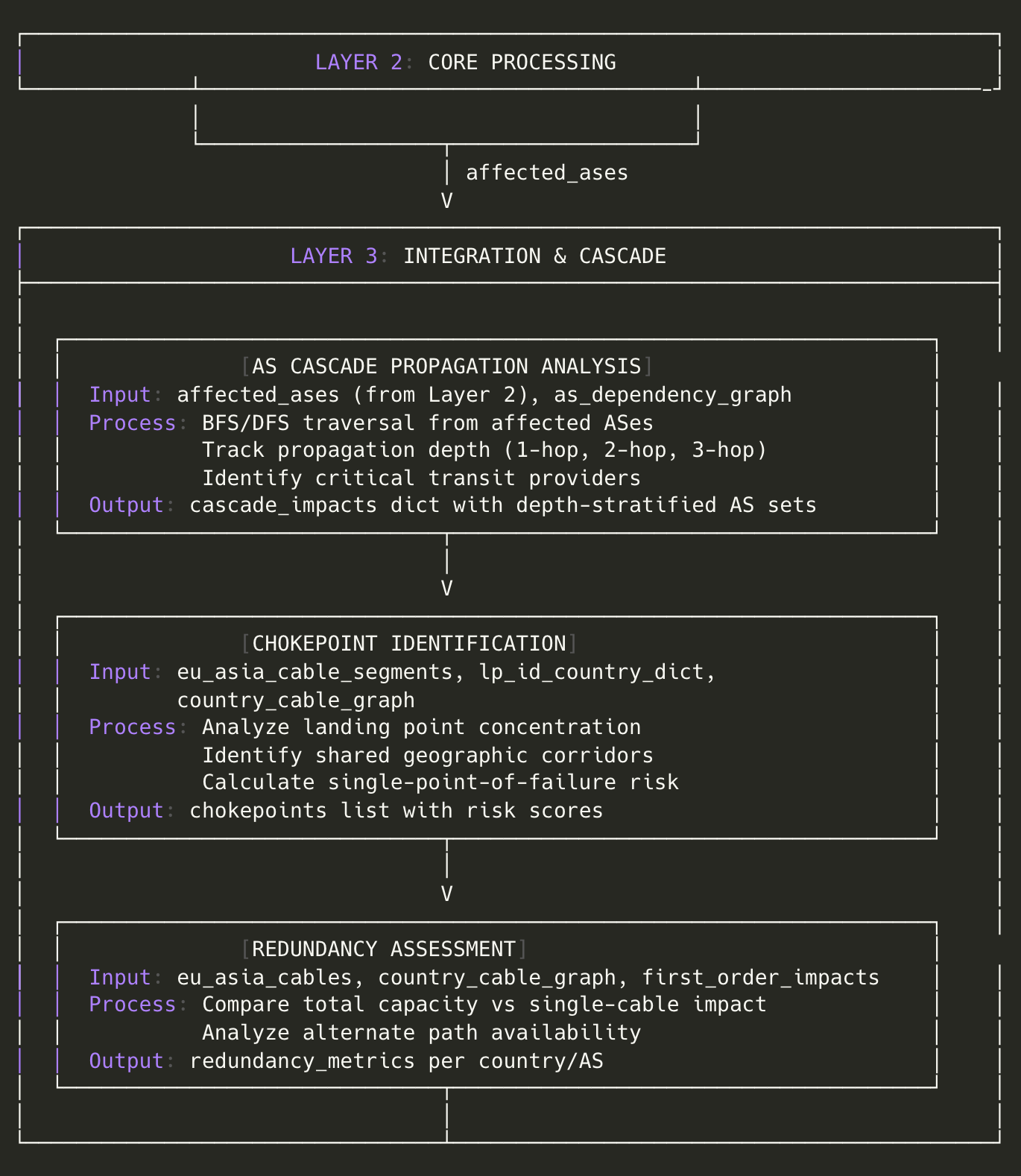}
\caption{Europe-Asia Cable Failure Cascade Analysis Workflow (Integration \& Cascade Layer)}
\label{fig:cascade_workflow3}
\end{figure*}

\begin{figure*}[p]
\centering
\includegraphics[width=0.8\textwidth]{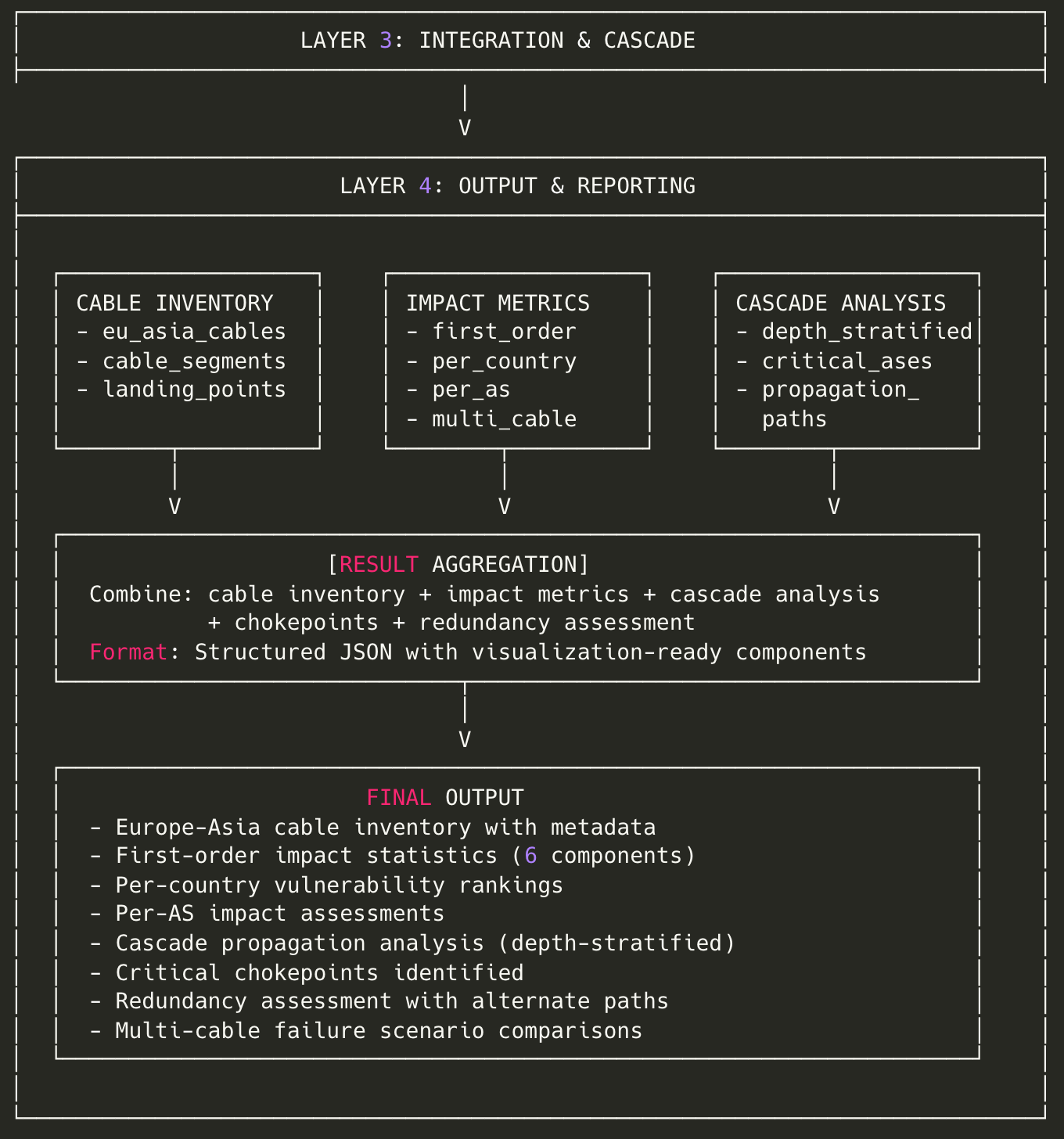}
\caption{Europe-Asia Cable Failure Cascade Analysis Workflow (Output Reporting Layer)}
\label{fig:cascade_workflow4}
\end{figure*}

\section{Workflow Composition Patterns}
\label{sec:patterns}
 WorkflowScout employs four method composition patterns to eliminate the need for manual parameter calibration while maintaining workflow robustness. The MERGE pattern combines multiple measurement methods algorithmically—for example, running two geolocation APIs and merging results through intersection (for high confidence) or union (for coverage)—thereby achieving robustness through methodological diversity. The AUTO\_CALIBRATE pattern derives parameters automatically from dataset distribution statistics rather than requiring manual tuning, such as calculating thresholds based on the data's statistical properties. The DERIVED pattern extracts parameter values from registry specifications or problem requirements (e.g., timeout values from registry specs, scale from problem constraints), ensuring compatibility and feasibility. Finally, the CONSERVATIVE\_DEFAULT pattern applies loose bounds with explicitly documented trade-offs, such as using wide thresholds while documenting their precision-recall implications, which prevents premature data filtering while making the trade-offs transparent. Together, these patterns enable Airavat to generate measurement workflows that avoid the brittle manual parameter tuning typical of traditional approaches while maintaining methodological soundness.

\section{Sample Inputs and Outputs}

\begin{itemize}
    \item Figures~\ref{fig:kg_output} and \ref{fig:kg_output_2} show a sample knowledge graph output for a single paper.
    \item Figure~\ref{fig:querymind_schema} shows the QueryMind Schema.
    \item Table~\ref{tab:registry_functions} shows the registry functions used by Airavat's automated workflows across the various case studies.
    \item Figures~\ref{fig:cascade_workflow1}, ~\ref{fig:cascade_workflow2}, ~\ref{fig:cascade_workflow3}, ~\ref{fig:cascade_workflow4} show the output workflow generated for the cascading failure analysis.
\end{itemize}

\section{Architecture Details}
Table~\ref{tab:parameters} shows the Verification Engine parameters.

\end{document}